\documentclass[aps,pra,twocolumn,showkeys,secnumarabic]{revtex4-2}

\usepackage{physics}
\usepackage{tikz}
\usepackage{hyperref}
\usepackage{tikz-3dplot}
\usetikzlibrary{positioning}
\usetikzlibrary{decorations.pathmorphing}
\definecolor{darkgreen}{rgb}{0.0, 0.8, 0.2}

\usepackage{amssymb}
\usepackage{graphicx}%
\usepackage{multirow}%
\usepackage{amsmath,amssymb,amsfonts}%
\usepackage{amsthm}%
\usepackage{mathrsfs}%
\usepackage[title]{appendix}%
\usepackage{xcolor}%
\usepackage{textcomp}%

\usepackage{natbib}
\usepackage{algpseudocode}%
\usepackage{listings}%
\usetikzlibrary{shapes.geometric}
\usepackage{subcaption}
\usepackage{diagbox}
\def\commutator#1#2{[#1, #2]}
\usetikzlibrary{decorations.pathreplacing}

\begin{document}


\title{Polarization-encoded relativistic quantum key distribution}


\author{Georgi Bebrov}
\email[]{g.bebrov@tu-varna.bg}
\affiliation{Technical University of Varna, Bulgaria}


\date{\today}

\begin{abstract}
We propose a polarization-encoded relativistic quantum
key distribution (RQKD) protocol employing single-photon
states prepared in two nonorthogonal polarizations.
The proposed protocol exploits a polarization-dependent
measurement strategy that renders the probability of
a conclusive detection event independent of fluctuations
in the relative optical phase.
A relativistic security analysis is carried out to examine
the constraints imposed by relativistic causality on
an intercept-and-resend attack.
Furthermore, we investigate the influence of imperfect
polarization discrimination and background-induced
detection events on the quantum bit error rate and
consider a means of mitigating polarization leakage.
The secret-key rate of the proposed protocol is evaluated
over a turbulent free-space optical channel and compared
with that of standard phase-encoded RQKD.
The numerical results indicate that polarization encoding
can improve both the achievable secret-key rate and
the maximum transmission distance under the channel
conditions considered.
\end{abstract}


\maketitle

\section{Introduction}
\label{intro}
\indent Quantum key distribution (QKD) enables two distant
parties to establish a shared secret key by exploiting
the fundamental principles of quantum mechanics
\cite{Bennett1984,b92,ekert}.
Although QKD offers the possibility of information-theoretic
security, its practical realization remains subject to
limitations imposed by optical losses, detector imperfections,
and environmental disturbances. These effects become
particularly important in long-distance optical communication,
where they limit the achievable secret-key rate and
transmission distance.\\
\indent Over the past decades, considerable research efforts have
been devoted to overcoming these limitations, leading to
the development of various QKD protocols and techniques.
Among the most significant advances are the decoy-state
method, which enables secure QKD with weak coherent
states in the presence of photon-number-splitting attacks
\cite{hwang,ma2005,wang2005,wang2}, and measurement-device-independent
(MDI) QKD, which eliminates security vulnerabilities
associated with the measurement devices~\cite{Lo2012}.
More recently, twin-field (TF) QKD~\cite{lucamarini}
and its variants, including the sending-or-not-sending
(SNS) TF-QKD protocol~\cite{sns}, have opened up new
possibilities for extending the transmission distance
by overcoming the linear rate-loss scaling characteristic
of conventional repeaterless QKD protocols. While these developments primarily address the limitations
associated with practical quantum-state preparation,
measurement, and transmission, alternative approaches
to QKD also exploit physical constraints arising from
the space-time arrangement of the communication process.
\indent Relativistic quantum key distribution (RQKD)
\cite{kravtsov,molotkov,kulik,bebrov,bebrov2}
provides an alternative framework in which quantum
information processing is supplemented by constraints
arising from relativistic causality.
The underlying communication mechanism consists of
\textit{space-time spreading} of a quantum state at the
transmitter and its subsequent \textit{space-time localization}
at the receiver~\cite{kravtsov,molotkov}.
The practical realization of RQKD, however, requires
careful consideration of the optical encoding and
detection mechanisms, particularly in the presence
of channel disturbances.\\
\indent Phase-encoded implementations of RQKD are inherently
sensitive to fluctuations in the relative optical phase.
In free-space optical (FSO) channels, atmospheric turbulence
can modify the interference conditions at the receiver,
giving rise to erroneous detection events and, consequently,
an increased quantum bit error rate (QBER).
Polarization encoding represents a natural alternative,
since atmospheric turbulence induces only minor
polarization distortions~\cite{toyoshima}.
Nevertheless, polarization encoding alone does not
necessarily eliminate phase-induced detection errors
in an interference-based receiver.\\
\indent In this work, we propose a polarization-encoded
RQKD protocol employing single-photon states prepared
in two nonorthogonal polarizations, namely horizontal
($H$) and diagonal ($D$). The principal modification consists of adapting the
polarization measurement basis to the polarization
selected by the receiving party.
Specifically, the horizontal--vertical ($H/V$) and
diagonal--antidiagonal ($D/A$) bases are employed for
horizontal and diagonal polarization selections,
respectively.
In either case, conclusive detection events are associated
with the polarization orthogonal to that selected
by the receiver.
Under ideal polarization preparation and analysis,
this measurement strategy renders the conclusive
detection probability independent of the relative
optical phase, thereby suppressing turbulence-induced
phase errors without requiring active phase compensation.\\
\indent Particular attention is devoted to the influence
of practical implementation imperfections on the performance
of the proposed protocol. In this regard, we examine the
contributions of imperfect polarization discrimination
and background-induced detection events to the quantum
bit error rate, as well as the possibility of mitigating
errors arising from imperfections of the optical system.\\
\indent The performance of the proposed protocol is evaluated
in terms of the achievable secret-key rate and transmission
distance over a turbulent FSO channel. A comparative
analysis with the standard phase-encoded RQKD protocol
is carried out to assess the potential advantages and
limitations of polarization encoding, particularly with
regard to its robustness against atmospheric disturbances.\\
\indent The paper is organized as follows.
Section~\ref{protocol} introduces the proposed polarization-encoded
RQKD protocol and describes its operational principles.
Section~\ref{sec-analysis} is devoted to the relativistic security
analysis of the proposed protocol.
Section~\ref{evaluation} addresses the evaluation of the secret-key rate,
while section \ref{num-results} presents the numerical results and a
comparative performance analysis of the polarization-encoded
and phase-encoded RQKD protocols.
Section~\ref{summary} summarizes the main findings and concludes
the paper.
The appendices provide detailed mathematical derivations
supplementing the analysis presented in the main text.

\section{Proposed protocol}
\label{protocol}
\begin{figure*}
	\centering
	\begin{tikzpicture}
		\draw (0,0) rectangle (0.3,0.3);
		\draw (0,0) -- (0.3,0.3);
		\node[above right] at (0.2,0.2) {BS$_{s}$};

		\draw (6,2) rectangle (6.3,2.3);
		\draw (6,2) -- (6.3,2.3);
		\node[below left] at (6.25,2.1) {BS$_{f}$}; 
		
		\draw (8,2) rectangle (8.3,2.3);
		\draw (8,2) -- (8.3,2.3);
		\node[below left] at (8.25,2.1) {PBS};
		
		\draw (6,4) rectangle (6.3,4.3);
		\draw (6,4.3) -- (6.3,4);
		\node[below left] at (7.15,4.1) {PBS};
		
		\draw[black] (0,2) -- (0.3,2.3);
		\draw[black] (6,0) -- (6.3,0.3);
		\node[left] at (0.05,2.15) {M};
		\node[right] at (6.15,0) {M};
		
		\draw[->,red] (-0.4,0.15) -- (0,0.15);
		\draw[->,red] (0.15,-0.2) -- (0.15,0);
		\node[fill=white,left] at (-0.4,0.15) {\fbox{$a^\dagger$}};
		\node[fill=white,below] at (0.15,-0.2) {\fbox{$b^\dagger$}$\;(\ket{0})$};
		\draw[red] (0.3,0.15) -- (6.15,0.15) -- (6.15,2);
		\node[fill=white,right] at (1,0.15) {\fbox{$l^\dagger$}};
		
		\draw[red] (0.15,0.3) -- (0.15,2.15) -- (6,2.15);
		\node[fill=white,above] at (0.15,1) {\fbox{$u^\dagger$}};
		
		\draw[fill=blue!40] (0.6,2.15) circle (0.15);
		\node[above] at (0.6,2.2) {$\hat{H}_a$};
		\draw[fill=blue!40] (6.15,1.15) circle (0.15);
		\node[right] at (6.2,1.15) {$\hat{H}_b$};
		
		\draw[->,red] (6.15,2.3) -- (6.15,4.7);
		\draw[->,red] (6.3,2.15) -- (8.7,2.15);
		\node[fill=white,right] at (6.7,2.15) {\fbox{$d^\dagger$}};
		\node[fill=white,above] at (6.15,2.7) {\fbox{$c^\dagger$}};

		\draw[->,red] (6.15,4.15) -- (5.3,4.15);
		\draw (5.2,3.9) arc[start angle=-90,end angle=-270,radius=0.25];
		\node[right] at (4.85,3.65) {$V^{\text{\sffamily \textbf{u}}}$ ($A^{\text{\sffamily \textbf{u}}}$)};

		\draw (6.4,4.75) arc[start angle=0,end angle=180,radius=0.25];
		\node[right] at (6.45,4.75) {$H^{\text{\sffamily \textbf{u}}}$ ($D^{\text{\sffamily \textbf{u}}}$)};

		\draw[->,red] (8.125,2.15) -- (8.125,2.95);
		\draw (8.8,2.4) arc[start angle=90,end angle=-90,radius=0.25];
		\node[right] at (9,2.15) {D};
		\node[right] at (8.4,1.7) {$H^{\text{\sffamily \textbf{l}}}$ ($D^{\text{\sffamily \textbf{l}}}$)};

		\draw (8.375,3) arc[start angle=0,end angle=180,radius=0.25];
		\node[right] at (7.6,3.45) {$V^{\text{\sffamily \textbf{l}}}$ ($A^{\text{\sffamily \textbf{l}}}$)};

		\draw[dashed,fill=green!50,fill opacity=0.2] (-1.25,-1) rectangle (1,3.75);
		
		\draw[dashed,fill=blue!50,fill opacity=0.2] (4.8,-1) rectangle (9.85,5.25);
		
		\node[] at (-0.125,3.5) {\underline{\textit{Alice}}};
		
		\node[] at (7,-0.75) {\underline{\textit{Bob}}};
		
		\node[] at (3.125,1.25) {\underline{\textit{quantum channel}}};

		\draw[<->] (4.5,2.15) -- (4.5,0.15);
		\node[] at (4.25,0.5) {$r$};
		
		\draw[-,color=teal,line width=1.25mm] (7.85,2.65) -- (8.4,2.65);
		\node[right,color=teal] at (8.4,2.65) {polarizer};

		\draw[-,color=teal,line width=1.25mm] (5.65,3.875) -- (5.65,4.4);
		
	\end{tikzpicture}
	\caption{\label{setup} Setup of a polarization-based RQKD in the single\textendash photon regime. BS\textemdash \textit{beam splitter}; PBS\textemdash \textit{polarizing beam splitter}; M\textemdash \textit{mirror}; D\textemdash \textit{detector}; $\hat{H}_{a(b)}$\textemdash \textit{Hadamard operator} applied by \textit{Alice} (\textit{Bob}); $H$\textemdash horizontal polarization detection; $V$\textemdash vertical polarization detection; $D$\textemdash diagonal polarization detection; $A$\textemdash off-diagonal (anti-diagonal) polarization detection; $r$\textemdash \textit{interferometer arms separation}; $\bullet^{\text{\sffamily \textbf{u}}}$\textemdash upper output detector; $\bullet^{\text{\sffamily \textbf{l}}}$\textemdash lower output detector. Note that a \textit{vacuum} state ($\ket{0}_b$) is fed into input $b$. The polarizers (polarization filter) are aligned with the polarization to be detected in the corresponding path ($V$ or $A$).}
\end{figure*}
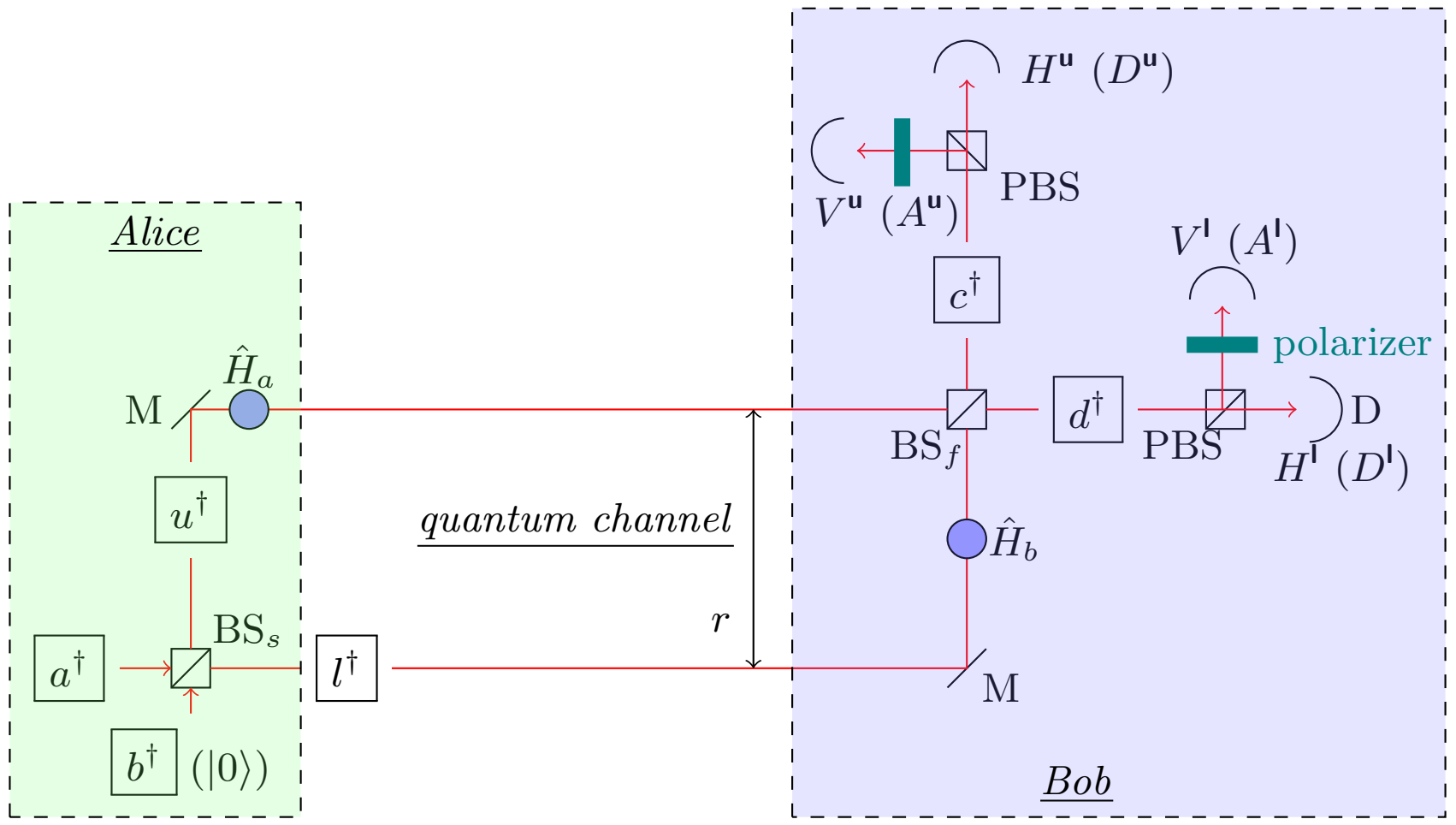

This work reports a polarization-encoding relativistic quantum key distribution protocol. It is characterized by the setup of figure \ref{setup}. The protocol consists of the following steps.\\
\textit{\textbf{Step}} 1. \textit{Alice} prepares $k$ ($k=0,1,2,...,N-1$) single photons in a horizontal polarization $H$. Here $N$ identifies the number of photons to be transferred from \textit{Alice} to \textit{Bob}. All the photons reside in horizontally-polarized optical mode $a^\dagger_{\text{H}}$.\\
\textit{\textbf{Step}} 2. At time $t_k = t_0 + kT$ ($t_0$\textemdash protocol initialization time instant; $T$\textemdash time interval between distinct $k$ rounds), the \textit{k}th photon is fed into the setup (interferometer) used in the protocol (see for reference figure \ref{setup}). Upon entering the interferometer, each photon passes through the beam splitter BS$_s$. Passing BS$_s$, the photon jumps over to a superposition $a^\dagger_{\text{H}}$ $\rightarrow$ $(u^\dagger_{\text{H}}+l^\dagger_{\text{H}})/\sqrt{2}$.  
At time instant $t_0 + \frac{r}{2c} + kT$, according to a probability distribution $\{p_a, 1-p_a\}$, \textit{Alice} applies an Hadamard operator $\hat{H}_a$ to the upper-arm mode of the interferometer: $u^\dagger_{\text{H}} \rightarrow u^\dagger_{\text{D}}$ if $\hat{H}_a$ is applied; otherwise $u^\dagger_{\text{H}}$ remains intact. The upper- and lower-arm modes travel through a free-space optical channel to the recipient, \textit{Bob}.\\
\textit{\textbf{Step}} 3. \indent At time $t'_k=t_0 + \frac{L}{c} + \frac{r/2}{c} + kT$ ($r$\textemdash interferometer's height/width; $L$\textemdash interferometer's length), according to a probability distribution $\{p_b, 1-p_b\}$, \textit{Bob} applies an Hadamard operator $\hat{H}_b$ to the lower-arm mode of the interferometer: $l^\dagger_{\text{H}} \rightarrow l^\dagger_{\text{D}}$ if $\hat{H}_b$ is applied; otherwise $l^\dagger_{\text{H}}$ remains intact.  \\
\textit{\textbf{Step}} 4. At time $t''_k = t_0 + \frac{L}{c} + \frac{r}{c} + kT$, the upper- and lower-arm modes interfere at the last beam splitter (BS$_f$) of the interferometer. The possible interference results are introduced in table \ref{interference}. Detection systems are placed only along each output, as shown in figure \ref{setup}. The detection system consists of a polarizing beam splitter (PBS) and detectors placed at its outputs. \textit{Bob} divides the $N$ rounds into two groups according to the probability distribution $\{p_r,1-p_r\}$: \textit{key-generation rounds}, selected with probability $p_r$, and \textit{test rounds}, selected with probability $1-p_r$. \\
\textit{Key-generation rounds}: \\
The PBSs are configured to measure in the \textit{rectilinear basis} $\{H,V\}$ if $\hat{H}_b$ is not applied, and in the \textit{diagonal basis }$\{D,A\}$ otherwise. \\
\textit{Test rounds}: \\
\textit{Bob} randomly selects one of three equally probable scenarios: \textbf{1)} the PBSs are configured to measure exclusively in the \textit{rectilinear basis} $\{H,V\}$, independently of \textit{Bob}'s choice of whether to apply $\hat{H}_b$; \textbf{2)} the PBSs are configured to measure exclusively in the \textit{diagonal basis} $\{D,A\}$, again independently of this choice. \\
\textit{\textbf{Step}} 5. (Key-generation) \textbf{In this step we consider only the key-generation rounds}. Table \ref{detections} presents the detection patterns given the modes entering the final beam splitter. Table \ref{probabilities} presents the detection probabilities given the modes entering the final beam splitter.
\textit{Bob} considers only $V$ (or $A$) detections as \textit{conclusive}. Also all detection clicks taking place at a time instance later than $t'''_k$ (for each $k$) are considered as intercepted (untrusted). These delayed detection events are discarded by \textit{Bob}. Each conclusive and non-delayed detection event is assigned the bit value \(0\) when \(\hat{H}_b\) is not applied and \(1\) when \(\hat{H}_b\) is applied. \textit{Bob} announces the $k$ rounds in which conclusive, non-delayed detection events take place. Based on this information, \textit{Alice} compiles \textit{her} key bits as follows. For each conclusive, non-delayed detection event is assigned the bit value \(1\) when \(\hat{H}_a\) is not applied and \(0\) when \(\hat{H}_a\) is applied. Note that key bits are established only when either ($u^\dagger_{\text{D}}$,$l^\dagger_{\text{H}}$) or ($u^\dagger_{\text{H}}$,$l^\dagger_{\text{D}}$) configuration takes place.\\
\textit{\textbf{Step}} 6. (Testing) \textbf{In this step only the test rounds are considered}. \textit{Alice} and \textit{Bob} use the detections in all testing rounds (scenarios \textbf{1)} and \textbf{2)}, see section \ref{sec-analysis}) for error-rate evaluation purposes (and also detecting the presence of an intercept-and-resend attack). If the error rate exceeds a preliminary determined threshold, the current protocol session is terminated. Otherwise, \textit{they} proceed to the next step. \\
\textit{\textbf{Step}} 7. \textit{Alice} and \textit{Bob} perform an information reconciliation (error correction) procedure.\\
\textit{\textbf{Step}} 8. \textit{Alice} and \textit{Bob} perform a privacy amplification procedure.\\
\fbox{\textit{\color{red}Note}:} For the sake of symmetry, we pick $p_a = p_b = 0.5$. \\
\indent {\color{red} For the sake of clarity, appendix \ref{math} presents a detailed mathematical formulation of the quantum stage of the proposed protocol.}

\begin{table}[!hbt] 
	\caption{\label{interference} All possible interference scenarios at BS$_f$. Note that the state $-\frac{1}{\sqrt{2}}u^\dagger_{\text{\sffamily X}}-\frac{1}{\sqrt{2}}l^\dagger_{\text{\sffamily Y}}$ is fed into the final beam splitter. Notice that $\text{\sffamily X}$, $\text{\sffamily Y}$ $\in\{H,D\}$. Take into account that the minus signs (phases of $\pi$) of both modes are provided by the reflections off the mirrors. In the table, no relative phase drift between $u^\dagger$ and $l^\dagger$ is accounted for.}
	\begin{ruledtabular}
		\begin{tabular}{ccc}
			\textbf{upper arm} & \textbf{lower arm} & \textbf{interference result} \\
			\hline
			$u^\dagger_{\text{H}}$ & $l^\dagger_{\text{H}}$ & $c^\dagger_{\text{H}}$\\
			$u^\dagger_{\text{H}}$ & $l^\dagger_{\text{D}}$ & $\frac{1}{2}(c^\dagger_{\text{H}}+c^\dagger_{\text{D}}+d^\dagger_{\text{H}}-d^\dagger_{\text{D}})$ \\
			$u^\dagger_{\text{D}}$ & $l^\dagger_{\text{H}}$ & $\frac{1}{2}(c^\dagger_{\text{D}}+c^\dagger_{\text{H}}+d^\dagger_{\text{D}}-d^\dagger_{\text{H}})$\\
			$u^\dagger_{\text{D}}$ & $l^\dagger_{\text{D}}$ & $c^\dagger_{\text{D}}$\\
		\end{tabular}
	\end{ruledtabular}
\end{table}

\begin{table}[!hbt] 
	\caption{\label{detections} Detection patterns given upper- and lower-arm modes at BS$_f$ in key-generation rounds. In the table, no relative phase drift between $u^\dagger$ and $l^\dagger$ is accounted for. Furthermore no dark counts are considered and the detectors' efficiencies are assumed to be unity. Note that $\lor$ identifies 'logical \textbf{or}' and $\text{\sffamily Z}^{\text{\sffamily \textbf{z}}}$ corresponds to a click at $\text{\sffamily Z}$ detector ($\text{\sffamily Z}\in\{H,V,D,A\}$) located along $\text{\sffamily z}$ output ($\text{\sffamily z}\in\{\text{\sffamily u},\text{\sffamily l}\}$, upper / lower output).}
	\begin{ruledtabular}
		\begin{tabular}{cccc}
			\textbf{upper arm} & \textbf{lower arm} & \textbf{basis} & \textbf{detection patterns} \\
			\hline
			$u^\dagger_{\text{H}}$ & $l^\dagger_{\text{H}}$ & $\{H,V\}$ &  $H^{\text{\sffamily \textbf{u}}}$ click\\
			$u^\dagger_{\text{H}}$ & $l^\dagger_{\text{D}}$ & $\{D,A\}$ & $D^{\text{\sffamily \textbf{u}}}$$\lor$$A^{\text{\sffamily \textbf{u}}}$$\lor$$D^{\text{\sffamily \textbf{l}}}$$\lor$$A^{\text{\sffamily \textbf{l}}}$ click \\
			$u^\dagger_{\text{D}}$ & $l^\dagger_{\text{H}}$ & $\{H,V\}$ & $D^{\text{\sffamily \textbf{u}}}$$\lor$$A^{\text{\sffamily \textbf{u}}}$$\lor$$D^{\text{\sffamily \textbf{l}}}$$\lor$$A^{\text{\sffamily \textbf{l}}}$ click\\
			$u^\dagger_{\text{D}}$ & $l^\dagger_{\text{D}}$ & $\{D,A\}$ & $D^{\text{\sffamily \textbf{u}}}$ click \\
		\end{tabular}
	\end{ruledtabular}
\end{table}

\begin{table}[!hbt] 
	\caption{\label{probabilities} Detection probabilities given upper- and lower-arm modes at BS$_f$ in key-generation rounds. In the table, no relative phase drift between $u^\dagger$ and $l^\dagger$ is accounted for. Furthermore no dark counts are considered and the detectors' efficiencies are assumed to be unity. Recall that $p_a = p_b = 1/2$.}
	\begin{ruledtabular}
		\begin{tabular}{cc}
			\textbf{modes}&\textbf{detection probabilities} \\
			\hline
			($u^\dagger_{\text{H}}$,$l^\dagger_{\text{H}}$)&$H^{\text{\sffamily \textbf{u}}}$: $25\%$\\
			($u^\dagger_{\text{H}}$,$l^\dagger_{\text{D}}$)&$D^{\text{\sffamily \textbf{u}}}$: $18.21\%$,$A^{\text{\sffamily \textbf{u}}}$: $3.125\%$,$D^{\text{\sffamily \textbf{l}}}$: $0.54\%$,$A^{\text{\sffamily \textbf{l}}}$: $3.125\%$ \\
			($u^\dagger_{\text{D}}$,$l^\dagger_{\text{H}}$)&$D^{\text{\sffamily \textbf{u}}}$: $18.21\%$,$A^{\text{\sffamily \textbf{u}}}$: $3.125\%$,$D^{\text{\sffamily \textbf{l}}}$: $0.54\%$,$A^{\text{\sffamily \textbf{l}}}$: $3.125\%$\\
			($u^\dagger_{\text{D}}$,$l^\dagger_{\text{D}}$)&$D^{\text{\sffamily \textbf{u}}}$: $25\%$ \\
		\end{tabular}
	\end{ruledtabular}
\end{table}

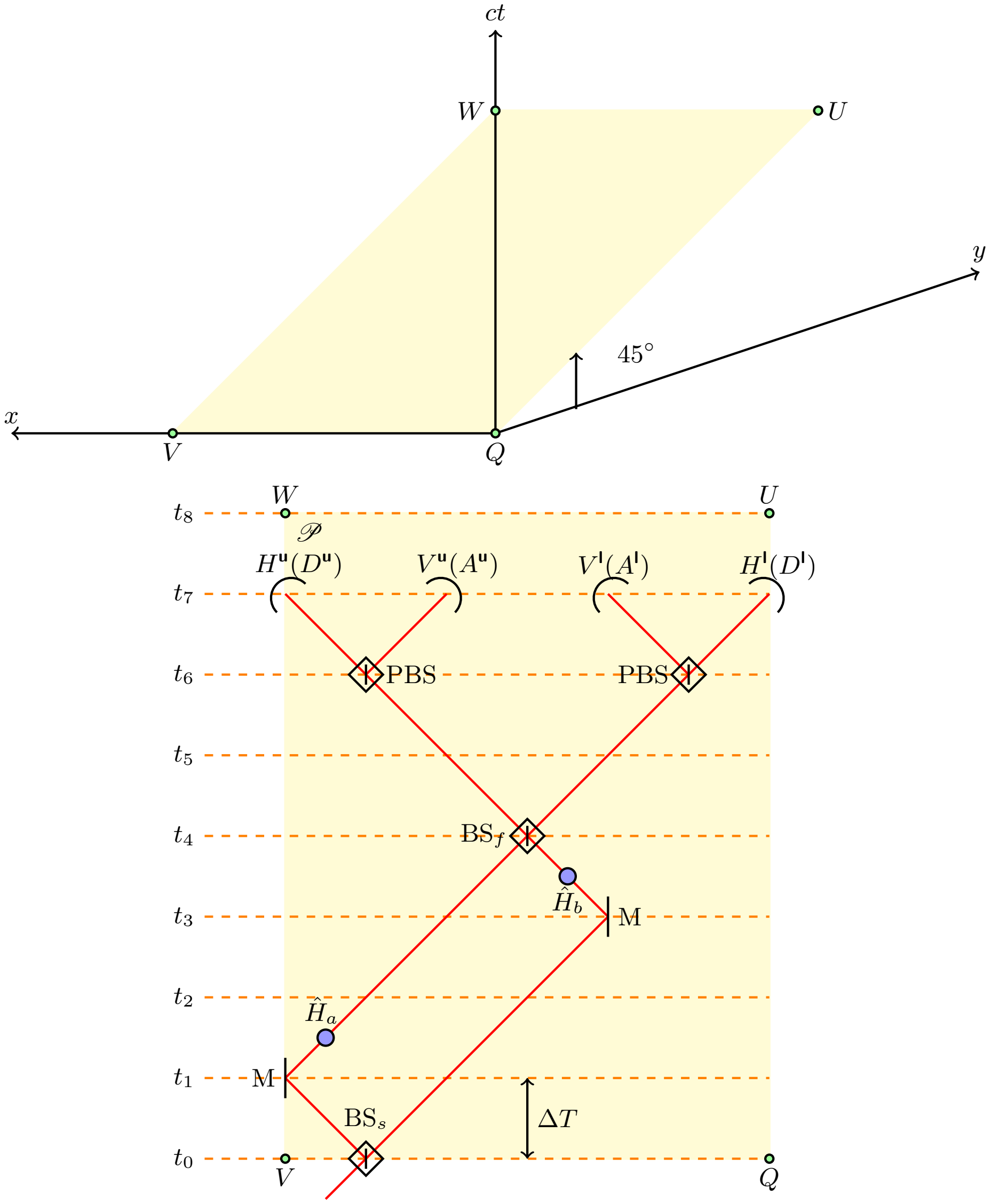
\begin{figure*}[!hbt]
	\centering
	\begin{tikzpicture}[thick]
		
		\draw[color=yellow!20,fill=yellow!20] (0,0) -- (4,4) -- (8,4) -- (4,0) -- cycle;
		\draw[->] (4,0) -- (-2,0) node[above] {$x$};
		\draw[->] (4,0) -- (4,5) node[above] {$ct$};
		\draw[->] (4,0) -- (10,2) node[above] {$y$};
		\node[] at (5.75,1) {$45^\circ$};
		\draw[->] (5,0.3) -| (5,1);
		
		\draw[fill=green!40] (0,0) circle (0.05);
		\draw[fill=green!40] (4,4) circle (0.05);
		\draw[fill=green!40] (8,4) circle (0.05);
		\draw[fill=green!40] (4,0) circle (0.05);
		
		\node[below] at (0,0) {$V$};
		\node[left] at (4,4) {$W$};
		\node[right] at (8,4) {$U$};
		\node[below] at (4,0) {$Q$};

	\end{tikzpicture}
	
	\begin{tikzpicture}[thick]
		
		\draw[color=yellow!20, fill=yellow!20] (0,0) rectangle (6,8);
		
		\draw[dashed,color=orange] (-1,0) -- (6,0);
		\node[left] at (-1,0) {$t_0$};
		
		\draw[dashed,color=orange] (-1,1) -- (6,1);
		\node[left] at (-1,1) {$t_1$};
		
		\draw[dashed,color=orange] (-1,2) -- (6,2);
		\node[left] at (-1,2) {$t_2$};
		
		\draw[dashed,color=orange] (-1,3) -- (6,3);
		\node[left] at (-1,3) {$t_3$};
		
		\draw[dashed,color=orange] (-1,4) -- (6,4);
		\node[left] at (-1,4) {$t_4$};
		
		\draw[dashed,color=orange] (-1,5) -- (6,5);
		\node[left] at (-1,5) {$t_5$};
		
		\draw[dashed,color=orange] (-1,6) -- (6,6);
		\node[left] at (-1,6) {$t_6$};
		
		\draw[dashed,color=orange] (-1,7) -- (6,7);
		\node[left] at (-1,7) {$t_7$};
		
		\draw[dashed,color=orange] (-1,8) -- (6,8);
		\node[left] at (-1,8) {$t_8$};
		
		\draw[fill=green!40] (0,0) circle (0.05);
		\draw[fill=green!40] (0,8) circle (0.05);
		\draw[fill=green!40] (6,8) circle (0.05);
		\draw[fill=green!40] (6,0) circle (0.05);
		
		\node[below] at (0,0) {$V$};
		\node[above] at (0,8) {$W$};
		\node[below right] at (0,8) {$\mathscr{P}$};
		\node[above] at (6,8) {$U$};
		\node[below] at (6,0) {$Q$};
		
		
		\draw[color=red] (0.5,-0.5) -- (4,3);
		\draw[color=red] (4,3) -- (0,7);
		
		\draw[color=red] (1,0) -- (0,1);
		\draw[color=red] (0,1) -- (6,7);
		
		\draw[color=red] (5,6) -- (4,7);
		\draw[color=red] (1,6) -- (2,7);

		\draw (4,2.75) -- (4,3.25);
		\node[right] at (4,3) {M};
		\draw (0,0.75) -- (0,1.25);
		\node[left] at (0,1) {M};
		
		\node[
		draw,
		diamond,
		minimum width=1,
		minimum height=0.5] at (3,4) {};
		\draw (3,3.875) -- (3,4.125);
		\node[left] at (2.875,4) {BS$_f$};
		
		\node[
		draw,
		diamond,
		minimum width=1,
		minimum height=0.5] at (5,6) {};
		\draw (5,5.875) -- (5,6.125);
		\node[left] at (4.875,6) {PBS};
		
		\node[
		draw,
		diamond,
		minimum width=1,
		minimum height=0.5] at (1,6) {};
		\draw (1,5.875) -- (1,6.125);
		\node[left] at (2,6) {PBS};

		
		\node[
		draw,
		diamond,
		minimum width=1,
		minimum height=0.5] at (1,0) {};
		\draw (1,-0.125) -- (1,0.125);
		\node[above] at (1,0.25) {BS$_s$};
		
		\draw[fill=blue!40] (0.5,1.5) circle (0.1);
		\node[above] at (0.45,1.55) {$\hat{H}_a$};
		\draw[fill=blue!40] (3.5,3.5) circle (0.1);
		\node[below] at (3.5,3.5) {$\hat{H}_b$};
		
		\draw (0.25,7.125) arc[start angle=45,end angle=225,radius=0.25];
		\node[right] at (-0.5,7.35) {$H^{\text{\sffamily \textbf{u}}} (D^{\text{\sffamily \textbf{u}}})$};
		
		\draw (1.75,7.125) arc[start angle=135,end angle=-45,radius=0.25];
		\node[right] at (1.5,7.35) {$V^{\text{\sffamily \textbf{u}}} (A^{\text{\sffamily \textbf{u}}})$};
		
		\draw (4.25,7.125) arc[start angle=45,end angle=225,radius=0.25];
		\node[right] at (3.5,7.35) {$V^{\text{\sffamily \textbf{l}}} (A^{\text{\sffamily \textbf{l}}})$};
		
		\draw (5.75,7.125) arc[start angle=135,end angle=-45,radius=0.25];
		\node[right] at (5.5,7.35) {$H^{\text{\sffamily \textbf{l}}} (D^{\text{\sffamily \textbf{l}}})$};

		\draw[<->] (3,0) -- (3,1);
		\node[right] at (3,0.5) {$\Delta T$};
		
	\end{tikzpicture}
	
	\caption{\label{spacetime-single} Spacetime diagram of the proposed RQKD setup. The setup is placed on $\mathscr{P}$ ($VWUQ$) plane. Note that the dashed lines represent planes of constant time $t=t_i$. The time interval $\Delta T$ is a function of the arms separation $r$ ($\Delta T \propto r$).}
\end{figure*}

\section{Relativistic security analysis}
\label{sec-analysis} 

 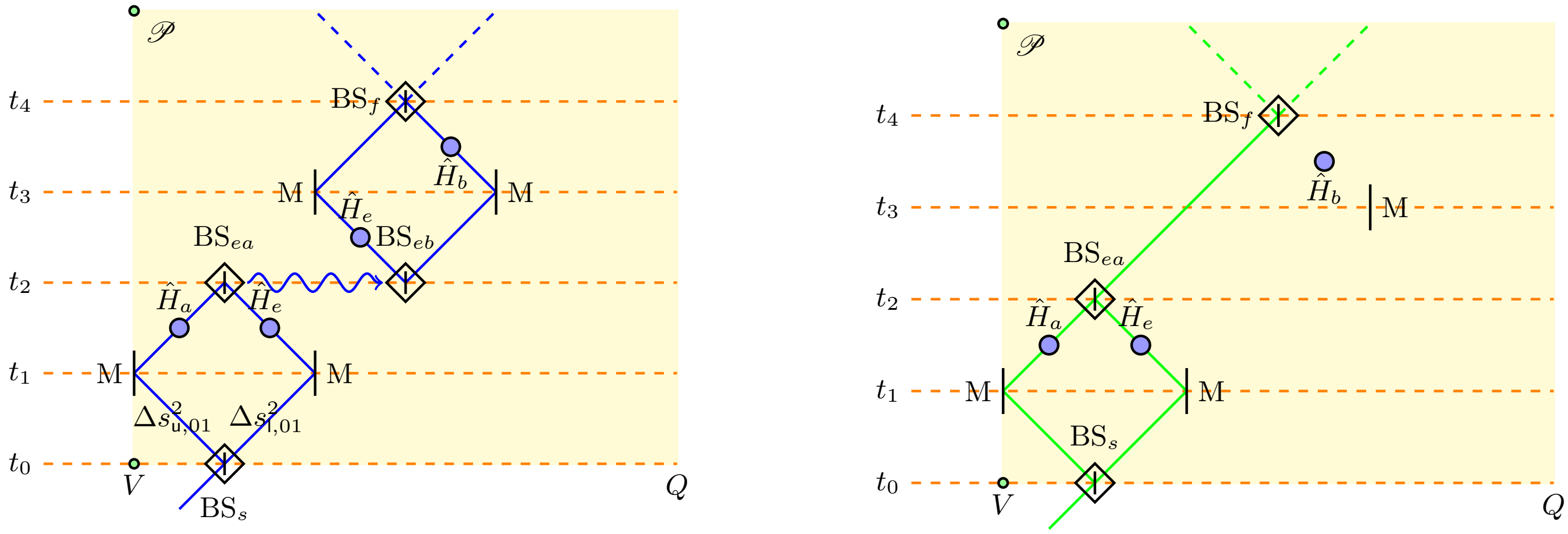
\begin{figure*}[!hbt]
 	\centering
 	
 	\begin{subfigure}[t]{0.45\textwidth}
 		\centering
 		\resizebox{\linewidth}{!}{%
 		
 	\begin{tikzpicture}[thick]
 		
 		\draw[color=yellow!20, fill=yellow!20] (0,0) rectangle (6,5);
 		
 		\draw[dashed,color=orange] (-1,0) -- (6,0);
 		\node[left] at (-1,0) {$t_0$};
 		
 		\draw[dashed,color=orange] (-1,1) -- (6,1);
 		\node[left] at (-1,1) {$t_1$};
 		
 		\draw[dashed,color=orange] (-1,2) -- (6,2);
 		\node[left] at (-1,2) {$t_2$};
 		
 		\draw[dashed,color=orange] (-1,3) -- (6,3);
 		\node[left] at (-1,3) {$t_3$};
 		
 		\draw[dashed,color=orange] (-1,4) -- (6,4);
 		\node[left] at (-1,4) {$t_4$};

 		\draw[fill=green!40] (0,0) circle (0.05);
 		\draw[fill=green!40] (0,5) circle (0.05);

 		\node[below] at (0,0) {$V$};
 		
 		\node[below right] at (0,5) {$\mathscr{P}$};
 	
 		\node[below] at (6,0) {$Q$};
 		
 		
 		\draw[color=blue] (0.5,-0.5) -- (2,1);
 		\draw[color=blue] (4,3) -- (3,4);
 		
 		\draw[color=blue] (1,0) -- (0,1);
 		\draw[color=blue] (0,1) -- (1,2);
 		\draw[color=blue] (2,1) -- (1,2);
 		
 		\draw[color=blue] (3,2) -- (2,3);
 		\draw[color=blue] (3,2) -- (4,3);
 		\draw[color=blue] (2,3) -- (3,4);
 		
 		\draw[color=blue,dashed] (3,4) -- (4,5);
 		\draw[color=blue,dashed] (3,4) -- (2,5);
 		
 		\draw (4,2.75) -- (4,3.25);
 		\node[right] at (4,3) {M};
 		\draw (0,0.75) -- (0,1.25);
 		\node[left] at (0,1) {M};
 		
 		\draw (2,2.75) -- (2,3.25);
 		\node[left] at (2,3) {M};
 		\draw (2,0.75) -- (2,1.25);
 		\node[right] at (2,1) {M};

 		\node[
 		draw,
 		diamond,
 		minimum width=1,
 		minimum height=0.5] at (3,4) {};
 		\draw (3,3.875) -- (3,4.125);
 		\node[left] at (2.875,4) {BS$_f$};

 		
 		\node[
 		draw,
 		diamond,
 		minimum width=1,
 		minimum height=0.5] at (1,0) {};
 		\draw (1,-0.125) -- (1,0.125);
 		\node[below] at (1,-0.25) {BS$_s$};

 		\node[
 		draw,
 		diamond,
 		minimum width=1,
 		minimum height=0.5] at (1,2) {};
 		\draw (1,1.875) -- (1,2.125);
 		\node[above] at (1,2.25) {BS$_{ea}$};
 		
 		\node[
 		draw,
 		diamond,
 		minimum width=1,
 		minimum height=0.5] at (3,2) {};
 		\draw (3,1.875) -- (3,2.125);
 		\node[above] at (3,2.25) {BS$_{eb}$};
 		
 		\draw[fill=blue!40] (0.5,1.5) circle (0.1);
 		\node[above] at (0.45,1.55) {$\hat{H}_a$};
 		\draw[fill=blue!40] (3.5,3.5) circle (0.1);
 		\node[below] at (3.5,3.5) {$\hat{H}_b$};
		\draw[fill=blue!40] (1.5,1.5) circle (0.1);
		\node[above] at (1.45,1.55) {$\hat{H}_e$};
		
 		\draw[fill=blue!40] (2.5,2.5) circle (0.1);
 		\node[above] at (2.45,2.55) {$\hat{H}_e$};
 		
 		\draw[->, decorate,
 		decoration={snake, amplitude=1mm, segment length=4mm},color=blue]
 		(1.25,2) -- (2.75,2);

 		\node[left, color=black] at (1,0.5) {$\Delta s^2_{\text{\sffamily u},01}$};
 		\node[left, color=black] at (2,0.5) {$\Delta s^2_{\text{\sffamily l},01}$};

 	\end{tikzpicture}
 		 		
 	}
 	\caption{Intercepted-and-resend attack: non-causal spacetime path of the intercepted photon.}
 	\label{spacetime-sec-a}
 	\end{subfigure}
	\hfill
	\begin{subfigure}[t]{0.45\textwidth}
		\centering
		\resizebox{\linewidth}{!}{%
			\begin{tikzpicture}[thick]
				
	\draw[color=yellow!20, fill=yellow!20] (0,0) rectangle (6,5);
	
	\draw[dashed,color=orange] (-1,0) -- (6,0);
	\node[left] at (-1,0) {$t_0$};
	
	\draw[dashed,color=orange] (-1,1) -- (6,1);
	\node[left] at (-1,1) {$t_1$};
	
	\draw[dashed,color=orange] (-1,2) -- (6,2);
	\node[left] at (-1,2) {$t_2$};
	
	\draw[dashed,color=orange] (-1,3) -- (6,3);
	\node[left] at (-1,3) {$t_3$};
	
	\draw[dashed,color=orange] (-1,4) -- (6,4);
	\node[left] at (-1,4) {$t_4$};

	\draw[fill=green!40] (0,0) circle (0.05);
	\draw[fill=green!40] (0,5) circle (0.05);

	\node[below] at (0,0) {$V$};

	\node[below right] at (0,5) {$\mathscr{P}$};

	\node[below] at (6,0) {$Q$};
	
	
	\draw[color=green] (0.5,-0.5) -- (2,1);
	
	\draw[color=green] (1,0) -- (0,1);
	\draw[color=green] (0,1) -- (1,2);
	\draw[color=green] (2,1) -- (1,2);
	
	\draw[color=green] (1,2) -- (3,4);

	\draw[color=green,dashed] (3,4) -- (4,5);
	\draw[color=green,dashed] (3,4) -- (2,5);
	
	\draw (4,2.75) -- (4,3.25);
	\node[right] at (4,3) {M};
	\draw (0,0.75) -- (0,1.25);
	\node[left] at (0,1) {M};

	\draw (2,0.75) -- (2,1.25);
	\node[right] at (2,1) {M};

	\node[
	draw,
	diamond,
	minimum width=1,
	minimum height=0.5] at (3,4) {};
	\draw (3,3.875) -- (3,4.125);
	\node[left] at (2.875,4) {BS$_f$};

	
	\node[
	draw,
	diamond,
	minimum width=1,
	minimum height=0.5] at (1,0) {};
	\draw (1,-0.125) -- (1,0.125);
	\node[above] at (1,0.25) {BS$_s$};

	\node[
	draw,
	diamond,
	minimum width=1,
	minimum height=0.5] at (1,2) {};
	\draw (1,1.875) -- (1,2.125);
	\node[above] at (1,2.25) {BS$_{ea}$};

	\draw[fill=blue!40] (0.5,1.5) circle (0.1);
	\node[above] at (0.45,1.55) {$\hat{H}_a$};
	\draw[fill=blue!40] (3.5,3.5) circle (0.1);
	\node[below] at (3.5,3.5) {$\hat{H}_b$};
		
	\draw[fill=blue!40] (1.5,1.5) circle (0.1);
	\node[above] at (1.45,1.55) {$\hat{H}_e$};

				
	\end{tikzpicture}%
		}
		\caption{Intercepted-and-resend attack: causal spacetime path of the intercepted photon.}
		\label{spacetime-sec-b}
	\end{subfigure}
 	\caption{\label{spacetime-sec} Spacetime security analysis of the proposed RQKD setup given an intercepted-send attack launched by \textit{Eve}. The setup is placed on $\mathscr{P}$ ($VWUQ$) plane, see figure \ref{spacetime-single} for reference. Note that the horizontal dashed lines represent planes of constant time $t=t_i$.}
 \end{figure*}

The standard (non-intercepted) spacetime path of a single photon has the following features:
\begin{enumerate}
	\item[\textbf{1}.] Each propagation segment of a single photon (upper-arm, lower-arm, upper-output, etc.) follows a \textit{lightlike} trajectory/path characterized by a vanishing spacetime interval, $\Delta s^2=0$. Recall that \cite{minkowski1915,taylor1992}
	\begin{equation*}
		\begin{split}
			\Delta s^2 = c^2\Delta t^2 - \Delta x^2 - \Delta y^2
		\end{split}
	\end{equation*}
	where $c$ denotes the \textit{speed of light}, $t$ is the time coordinate, whereas $x$ and $y$ are the spatial coordinates in the spacetime under consideration (see figure \ref{spacetime-single}).
	\item[\textbf{2}.] The upper- and lower-arm optical paths are arranged such that their propagation times are equal (see figure \ref{spacetime-single} for reference). This ensures that the corresponding optical modes interfere properly (overlap temporally) at BS$_f$.
\end{enumerate}
\indent We now examine the spacetime behavior of the proposed scheme under an \textit{intercept-and-resend} attack performed by an eavesdropper, \textit{Eve}. In this attack \textit{Eve} intercepts the single-photon quantum system transmitted by \textit{Alice} and sends a new quantum system to \textit{Bob}, prepared in a state conditioned on the outcome of her measurement of the intercepted system. The intercepted-and-resend attack is shown in figure \ref{spacetime-sec}. \\
\indent To avoid introducing errors in \textit{Bob}'s measurement outcomes (\textit{i.e.}, disturbing the interference process at BS$_f$), \textit{Eve} attempts to perform the intercepted-and-resend attack as illustrated in figure \ref{spacetime-sec-a}. In the following we demonstrate that such an attempt is incompatible with the principles of special relativity, as it would require a violation of causality. To demonstrate that the proposed attack violates causality, we analyze the spacetime intervals associated with the upper- and lower-arm optical paths. The upper-arm path is divided into segments characterized by $\Delta s^2_{\text{\sffamily u},01}$, $\Delta s^2_{\text{\sffamily u},12}$, $\Delta s^2_{\text{\sffamily u},22}$, $\Delta s^2_{\text{\sffamily u},23}$, and $\Delta s^2_{\text{\sffamily u},34}$, see figure \ref{spacetime-sec-a} for reference. Here the subscript $\text{\sffamily u}$ identifies the upper-arm path, whereas the indices $v$ and $w$ in $\Delta s^2_{\text{\sffamily u},vw}$ denote the initial and final time instants of the corresponding segment, respectively. Similarly, the lower-arm path is divided into segments characterized by $\Delta s^2_{\text{\sffamily l},01}$, $\Delta s^2_{\text{\sffamily l},12}$, $\Delta s^2_{\text{\sffamily l},22}$, $\Delta s^2_{\text{\sffamily l},23}$, and $\Delta s^2_{\text{\sffamily l},34}$.
For causality to be preserved, each segment along which information must propagate must satisfy
\begin{equation*}
	\begin{split}
		\Delta s^2_{j,vw}\geq 0,
		\qquad j\in\{\text{\sffamily u},\text{\sffamily l}\}.
	\end{split}
\end{equation*}
However the proposed attack requires information transfer across the segments characterized by
\begin{equation*}
	\begin{split}
		\Delta s^2_{\text{\sffamily u},22}<0,\qquad
		\Delta s^2_{\text{\sffamily l},22}<0,
	\end{split}
\end{equation*}
which are \textit{spacelike} intervals. In particular, since the initial and final events of each segment occur at the same coordinate time $t_2$, we obtain
\begin{equation*}
	\begin{split}
		\Delta s^2_{j,22}
		=-(\Delta x_{j,22})^2-(\Delta y_{j,22})^2<0,
		\qquad j\in\{\text{\sffamily u},\text{\sffamily l}\},
	\end{split}
\end{equation*}
provided that the corresponding events are spatially separated. Consequently, no causal signal can connect these events. Since the proposed intercept-and-resend strategy requires measurement-dependent information to propagate across these spacelike separations, it would violate causality and is therefore physically impossible to implement in the configuration illustrated in figure \ref{spacetime-sec-a}.\\
\indent As illustrated in figure \ref{spacetime-sec-b}, a causal intercept-and-resend strategy requires the single photon resent by \textit{Eve} to propagate directly toward BS$_f$. In this case \textit{Alice} and \textit{Bob} cannot establish correlated key bits, since the photon detected after BS$_f$ is unaffected by \textit{Bob}'s Hadamard operation $\hat{H}_b$. \\
\indent We now examine the detection statistics induced by an intercept-and-resend attack during the test rounds.
As discussed above, a causal implementation of the
considered attack requires the photon resent by
\textit{Eve} to propagate directly toward BS$_f$,
thereby bypassing \textit{Bob}'s Hadamard operation
$\hat{H}_b$.

Suppose that \textit{Eve} successfully identifies
\textit{Alice}'s polarization state and resends a
single photon in the corresponding state
$\ket{\text{\sffamily p}_a}$, where
$\ket{\text{\sffamily p}_0}=\ket{H}$ and
$\ket{\text{\sffamily p}_1}=\ket{D}$.
Although the resent photon depends only on
\textit{Alice}'s choice $a$, \textit{Alice} and
\textit{Bob} continue to associate each round
with their independently selected encoding
choices $(a,b)$.

Since the other input of BS$_f$ is in the vacuum
state, the output state becomes
\begin{equation*}
	\ket{\psi_E}
	=
	\frac{1}{\sqrt{2}}
	\left(
	c^\dagger_{\text{\sffamily p}_a}
	+
	d^\dagger_{\text{\sffamily p}_a}
	\right)\ket{0}.
\end{equation*}

Accordingly, the detection probability at the
lower output, resolved with respect to a
polarization state $\ket{\text{\sffamily q}}$,
is
\begin{equation*}
	P_E(\text{\sffamily q}^{\text{\sffamily l}}
	|a,b,m)
	=
	\frac12
	\left|
	\braket{\text{\sffamily q}}
	{\text{\sffamily p}_a}
	\right|^2,
\end{equation*}
where $m=0$ and $m=1$ correspond to the
rectilinear and diagonal measurement bases,
respectively. Notice that this probability
is independent of \textit{Bob}'s Hadamard
choice $b$.

Let $T$ denote the discrete random variable
representing \textit{Bob}'s choice of testing
configuration, with $T\in\{1,2\}$ and
$P(T=i)=1/2$, $i\in\{1,2\}$, conditional
on a test round. The events $T_i=\{T=i\}$
correspond to scenarios \textbf{1)} and
\textbf{2)}, respectively (see section \ref{protocol} for reference).

In the first test scenario (scenario \textbf{1)}),
the PBSs measure exclusively in the rectilinear
basis $\{H,V\}$. Using
$|\braket{H}{H}|^2=1$,
$|\braket{V}{H}|^2=0$, and
$|\braket{H}{D}|^2
=|\braket{V}{D}|^2=1/2$,
we obtain
\begin{equation*}
	\begin{split}
		P_E(H^{\text{\sffamily l}}|a,b,T_1)
		&=
		\begin{cases}
			1/2, & a=0,\\
			1/4, & a=1,
		\end{cases}\\
		P_E(V^{\text{\sffamily l}}|a,b,T_1)
		&=
		\begin{cases}
			0, & a=0,\\
			1/4, & a=1.
		\end{cases}
	\end{split}
\end{equation*}

In the second test scenario (scenario \textbf{2)}),
the PBSs measure exclusively in the diagonal
basis $\{D,A\}$. Since
$|\braket{D}{H}|^2
=|\braket{A}{H}|^2=1/2$,
$|\braket{D}{D}|^2=1$, and
$|\braket{A}{D}|^2=0$,
we similarly obtain
\begin{equation*}
	\begin{split}
		P_E(D^{\text{\sffamily l}}|a,b,T_2)
		&=
		\begin{cases}
			1/4, & a=0,\\
			1/2, & a=1,
		\end{cases}\\
		P_E(A^{\text{\sffamily l}}|a,b,T_2)
		&=
		\begin{cases}
			1/4, & a=0,\\
			0, & a=1.
		\end{cases}
	\end{split}
\end{equation*}

To identify the disturbance introduced by the
considered attack, \textit{Alice} and
\textit{Bob} restrict their analysis to test
rounds with identical encoding choices,
i.e., $a=b$. In legitimate operation, the
two polarization components interfere
constructively at the upper output of BS$_f$
and destructively at its lower output.
Consequently, in the absence of phase
fluctuations and for ideal optical components,
\begin{equation*}
	\begin{split}
		P(H^{\text{\sffamily l}}|a=b,T_1)
		&=
		P(V^{\text{\sffamily l}}|a=b,T_1)
		=0,\\
		P(D^{\text{\sffamily l}}|a=b,T_2)
		&=
		P(A^{\text{\sffamily l}}|a=b,T_2)
		=0.
	\end{split}
\end{equation*}

Thus, any lower-output detection in a round
with $a=b$ constitutes an erroneous event
in the ideal legitimate protocol. In contrast,
under the considered intercept-and-resend
attack, the photon enters only one input
of BS$_f$, and the destructive interference
at the lower output is absent.

In scenario \textbf{1)}, the relevant
configurations (for checking an eavesdropping) are $HH$ and $DD$.
For $HH$, we obtain
\begin{equation*}
	\begin{split}
		P_E(H^{\text{\sffamily l}}|HH,T_1)
		&=\frac12,\\
		P_E(V^{\text{\sffamily l}}|HH,T_1)
		&=0,
	\end{split}
\end{equation*}
whereas for $DD$,
\begin{equation*}
	\begin{split}
		P_E(H^{\text{\sffamily l}}|DD,T_1)
		&=\frac14,\\
		P_E(V^{\text{\sffamily l}}|DD,T_1)
		&=\frac14.
	\end{split}
\end{equation*}

Since both polarization detectors at the
lower output are monitored, the corresponding
erroneous-event probability is
\begin{equation*}
	\begin{split}
		p_{\text{error},E,T_1}
		&=
		\frac14
		\sum_{a=b}
		\Big[
		P_E(H^{\text{\sffamily l}}|a,b,T_1)\\
		&\qquad\qquad+
		P_E(V^{\text{\sffamily l}}|a,b,T_1)
		\Big]\\
		&=
		\frac14
		\left[
		\underbrace{\left(\frac12+0\right)}_{a=b=0}
		+
		\underbrace{\left(\frac14+\frac14\right)}_{a=b=1}
		\right]\\
		&=\frac14.
	\end{split}
\end{equation*}

Similarly, in scenario \textbf{2)}, the
$HH$ configuration yields
\begin{equation*}
	\begin{split}
		P_E(D^{\text{\sffamily l}}|HH,T_2)
		&=\frac14,\\
		P_E(A^{\text{\sffamily l}}|HH,T_2)
		&=\frac14,
	\end{split}
\end{equation*}
whereas the $DD$ configuration yields
\begin{equation*}
	\begin{split}
		P_E(D^{\text{\sffamily l}}|DD,T_2)
		&=\frac12,\\
		P_E(A^{\text{\sffamily l}}|DD,T_2)
		&=0.
	\end{split}
\end{equation*}

Consequently,
\begin{equation*}
	\begin{split}
		p_{\text{error},E,T_2}
		&=
		\frac14
		\sum_{a=b}
		\Big[
		P_E(D^{\text{\sffamily l}}|a,b,T_2)\\
		&\qquad\qquad+
		P_E(A^{\text{\sffamily l}}|a,b,T_2)
		\Big]\\
		&=
		\frac14
		\left[
		\left(\frac14+\frac14\right)
		+
		\left(\frac12+0\right)
		\right]\\
		&=\frac14.
	\end{split}
\end{equation*}

Since the two testing configurations are
selected with equal probabilities, the
overall erroneous-event probability,
conditioned on a test round, is
\begin{equation*}
	\begin{split}
		p_{\text{error},E,T}
		&=
		\frac12p_{\text{error},E,T_1}
		+
		\frac12p_{\text{error},E,T_2}\\
		&=\frac14.
	\end{split}
\end{equation*}

The above probability accounts for the
occurrence of identical encoding choices,
whose total probability is
$P(a=b)=1/2$. Therefore, the error
probability conditioned on $a=b$ is
\begin{equation*}
	\begin{split}
		P_E(\text{error}|a=b,T)
		&=
		\frac{p_{\text{error},E,T}}
		{P(a=b)}\\
		&=
		\frac{1/4}{1/2}
		=\frac12.
	\end{split}
\end{equation*}

In contrast, the corresponding probability
vanishes in the ideal legitimate protocol:
\begin{equation*}
	P(\text{error}|a=b,T)=0.
\end{equation*}

Therefore, the considered intercept-and-resend
attack produces a conditional error probability
of $50\%$ in the retained test rounds,
whereas the ideal legitimate protocol
exhibits no such errors. The disturbance
is revealed by detections at the lower
output of BS$_f$ for identical encoding
choices, irrespective of the measured
polarization component.

Notice that the resulting test statistic
is independent of the selected polarization
measurement basis, provided that both
polarization detectors at the lower output
are monitored. Hence, scenarios \textbf{1)}
and \textbf{2)} yield identical conditional
error probabilities for the considered attack.

\section{Key rate}
\label{evaluation}

\indent In the asymptotic regime, the secret-key rate of the
proposed polarization-encoded relativistic quantum key
distribution protocol can be expressed in the form~\cite{bebrov}
\begin{equation}
	R = Q_{\mathrm{pol}}
	\left[1-\chi-fH_2(e_{\mathrm{pol}})\right],
	\label{eq:key-rate-pol}
\end{equation}
where $Q_{\mathrm{pol}}$ denotes the conclusive single-photon
yield, $e_{\mathrm{pol}}$ is the quantum bit error rate (QBER),
$f\geq 1$ is the error correction inefficiency, $\chi$ quantifies
the information leakage to Eve, and $H_2(\bullet)$ is the binary
Shannon entropy, defined as
\begin{equation}
	H_2(x)=-x\log_2(x)-(1-x)\log_2(1-x).
\end{equation}
The quantities $\chi$ and $fH_2(e_{\mathrm{pol}})$ account
for the fractions of the raw key consumed in privacy
amplification and error correction, respectively.

\indent As established in appendix \ref{er-pol}, the conclusive
single-photon yield, taking into account channel losses,
detector inefficiency, imperfect polarization discrimination,
and background-induced detection events, is given by
\begin{equation}
	Q_{\mathrm{pol}}
	=
	Y_0^{\mathrm{pol}}
	+p_{\mathrm{det}}^{\mathrm{pol}}\eta,
	\label{eq:yield-pol-rate}
\end{equation}
where $\eta=\eta_t\eta_d$ denotes the product of the channel
transmissivity and detector efficiency, and
$Y_0^{\mathrm{pol}}$ is the background yield associated
with the conclusive detectors. For the cascaded polarization
analyzer considered in this work (PBS followed by a polarizer, see figure \ref{setup}), one has
\begin{equation}
	p_{\mathrm{det}}^{\mathrm{pol}}
	=\frac{t_1t_2+7\epsilon_1\epsilon_2}{8}.
\end{equation}
See appendix \ref{er-pol} for more details. Accordingly, equation~\eqref{eq:yield-pol-rate} takes the form
\begin{equation}
	Q_{\mathrm{pol}}
	=
	Y_0^{\mathrm{pol}}
	+\frac{t_1t_2+7\epsilon_1\epsilon_2}{8}\eta.
\end{equation}
The corresponding QBER is given by
equation~\eqref{eq:qber-pol}.

\indent In the ideal detection scenario, characterized by
$t_1=t_2=1$, $\epsilon_1=\epsilon_2=0$,
$\eta_d=1$, and $Y_0^{\mathrm{pol}}=0$, the conclusive
single-photon yield reduces to
\begin{equation}
	Q_{\mathrm{pol}}=\frac{\eta_t}{8}.
\end{equation}
Furthermore, in the absence of channel losses and other
sources of error, i.e., $\eta_t=1$ and
$e_{\mathrm{pol}}=0$, the secret-key rate reduces to
\begin{equation}
	R_{\mathrm{ideal}}=\frac{1}{8}(1-\chi).
	\label{eq:ideal-key-rate}
\end{equation}
Note that $Q_{\mathrm{pol}}$ is equal to  $p_{\text{det}}$ of appendix \ref{math} in this ideal scenario.

\subsection{Information leakage}

\indent As presented in section~\ref{protocol}, \textit{Alice} and \textit{Bob}
establish their raw key bits according to their respective
polarization choices, $H$ (horizontal polarization) and
$D$ (diagonal polarization). The two polarization states
are chosen independently and with equal a priori probabilities.\\
\indent We consider an eavesdropper, \textit{Eve}, who has access to the
quantum channel connecting \textit{Alice} and \textit{Bob}. 
Under these considerations, \textit{Alice}'s information source
is characterized by the ensemble
\begin{equation}
	\mathcal{S}
	=
	\left\{
	\frac{1}{2},\ket{H};
	\frac{1}{2},\ket{D}
	\right\},
\end{equation}
where
\begin{equation}
	\begin{split}
		\ket{H}
		&=
		\frac{1}{\sqrt{2}}
		\left(u_H^\dagger+l_H^\dagger\right)\ket{0},\\
		\ket{D}
		&=
		\frac{1}{\sqrt{2}}
		\left(u_D^\dagger+l_D^\dagger\right)\ket{0}.
	\end{split}
	\label{eq:source-states}
\end{equation}
Here, $u_p^\dagger$ and $l_p^\dagger$ denote the creation
operators associated with the upper and lower spatial modes,
respectively, for polarization $p\in\{H,D\}$.

\indent The density operator associated with the ensemble
$\mathcal{S}$ is given by
\begin{equation}
	\rho
	=
	\frac{1}{2}\ketbra{H}
	+\frac{1}{2}\ketbra{D}
	=
	\sum_{i=0}^{1}p_i\rho_i,
	\label{sigma}
\end{equation}
where $p_0=p_1=1/2$, $\rho_0=\ketbra{H}$, and
$\rho_1=\ketbra{D}$.

The accessible information associated with the ensemble
$\mathcal{S}$ is upper bounded by the Holevo
information~\cite{lima,lin-lutken}, given by
\begin{equation}
	\chi(\mathcal{S})
	=
	S(\rho)-\sum_{i=0}^{1}p_iS(\rho_i),
	\label{eq:holevo-pol}
\end{equation}
where $S(\bullet)$ denotes the von Neumann entropy.
Since both $\rho_0$ and $\rho_1$ are pure states
(see appendix~\ref{pure}), one has
$S(\rho_0)=S(\rho_1)=0$. Consequently,
\begin{equation}
	\chi(\mathcal{S})=S(\rho).
\end{equation}

\indent Taking into account the overlap between the two
single-photon states,
\begin{equation}
	\braket{H}{D}=\frac{1}{\sqrt{2}},
\end{equation}
the two nonzero eigenvalues of $\rho$ are obtained as
\begin{equation}
	\begin{split}
		\lambda_0
		&=
		\frac{1}{2}
		\left(1+\frac{1}{\sqrt{2}}\right),\\
		\lambda_1
		&=
		\frac{1}{2}
		\left(1-\frac{1}{\sqrt{2}}\right).
	\end{split}
\end{equation}
It thus follows that
\begin{equation}
	\begin{split}
		\chi(\mathcal{S})
		&=
		-\lambda_0\log_2(\lambda_0)
		-\lambda_1\log_2(\lambda_1)\\
		&=
		H_2\left[
		\frac{1}{2}
		\left(1+\frac{1}{\sqrt{2}}\right)
		\right]\\
		&\simeq 0.601.
	\end{split}
	\label{eq:holevo-value}
\end{equation}

\indent In the following rate-distance analysis, we adopt
the Holevo information of \textit{Alice}'s source ensemble as the
information leakage parameter, i.e.,
$\chi=\chi(\mathcal{S})$. This corresponds to an
information-leakage model in which \textit{Eve} is granted access
to the states transmitted by \textit{Alice}.
The resulting key rate is evaluated under this assumption.

\subsection{Key rate over a turbulent free-space optical channel}

\indent Since a turbulent free-space optical (FSO) channel
is considered, the channel transmissivity is expressed as
\begin{equation}
	\eta_t(L)=\eta_{\mathrm{lt}}(L)\eta_{\mathrm{atm}}(L),
\end{equation}
where $L$ denotes the transmission distance,
$\eta_{\mathrm{lt}}$ accounts for the finite receiver
aperture and long-term beam spreading, and
$\eta_{\mathrm{atm}}$ accounts for atmospheric extinction \cite{bebrov}. The overall transmissivity is thus given by
\begin{equation}
	\eta(L)
	=
	\eta_{\mathrm{lt}}(L)
	\eta_{\mathrm{atm}}(L)
	\eta_d.
	\label{eq:overall-transmissivity}
\end{equation}
The parameters characterizing the turbulent FSO channel
and the optical system are listed in
table~\ref{table-parameters}.\\
\indent Substituting the conclusive single-photon yield
into equation~\eqref{eq:key-rate-pol}, the secret-key rate of
the proposed protocol as a function of transmission
distance is expressed as
\begin{equation}
	\begin{split}
		R_{\mathrm{pol}}(L)
		=
		\max\Bigg\{0,\,
		&\left[
		Y_0^{\mathrm{pol}}
		+\frac{t_1t_2+7\epsilon_1\epsilon_2}{8}
		\eta(L)
		\right]\\
		&\times
		\left[
		1-\chi-fH_2(e_{\mathrm{pol}}(L))
		\right]
		\Bigg\},
	\end{split}
	\label{rate-loss}
\end{equation}
where the distance-dependent QBER is given by
\begin{equation}
	e_{\mathrm{pol}}(L)
	=
	\frac{
		e_0Y_0^{\mathrm{pol}}
		+\frac{1}{2}\epsilon_1\epsilon_2\eta(L)
	}{
		Y_0^{\mathrm{pol}}
		+\frac{1}{8}
		(t_1t_2+7\epsilon_1\epsilon_2)\eta(L)
	}.
	\label{eq:distance-qber-pol}
\end{equation}

\indent Although polarization distortions induced by
atmospheric turbulence are neglected, the secret-key rate
is still affected by turbulence through the reduction in
channel transmissivity. As the transmission distance
increases, the probability of signal-induced conclusive
detection events decreases, thereby increasing the relative
contribution of background-induced errors to the QBER.
Consequently, the maximum achievable transmission distance
is ultimately limited by the background yield, atmospheric
extinction, beam spreading, and detector efficiency, as
well as by the residual imperfections of the polarization
analyzer (PBS followed by a polarizer, see figure \ref{setup}).

\section{Numerical results}
\label{num-results}
\indent In this section, we evaluate the secret-key rate of
the proposed polarization-encoded RQKD protocol
(section~\ref{protocol}) as a function of transmission distance
over a turbulent free-space optical (FSO) channel.
The parameters characterizing the FSO channel are listed
in table~\ref{table-parameters} and are adopted from
Ref.~\cite{bebrov}. As discussed previously, polarization
distortions induced by atmospheric turbulence are neglected
throughout this work~\cite{toyoshima}.

\begin{figure*}[!htb]
	\centering
	\includegraphics[scale=0.6]
	{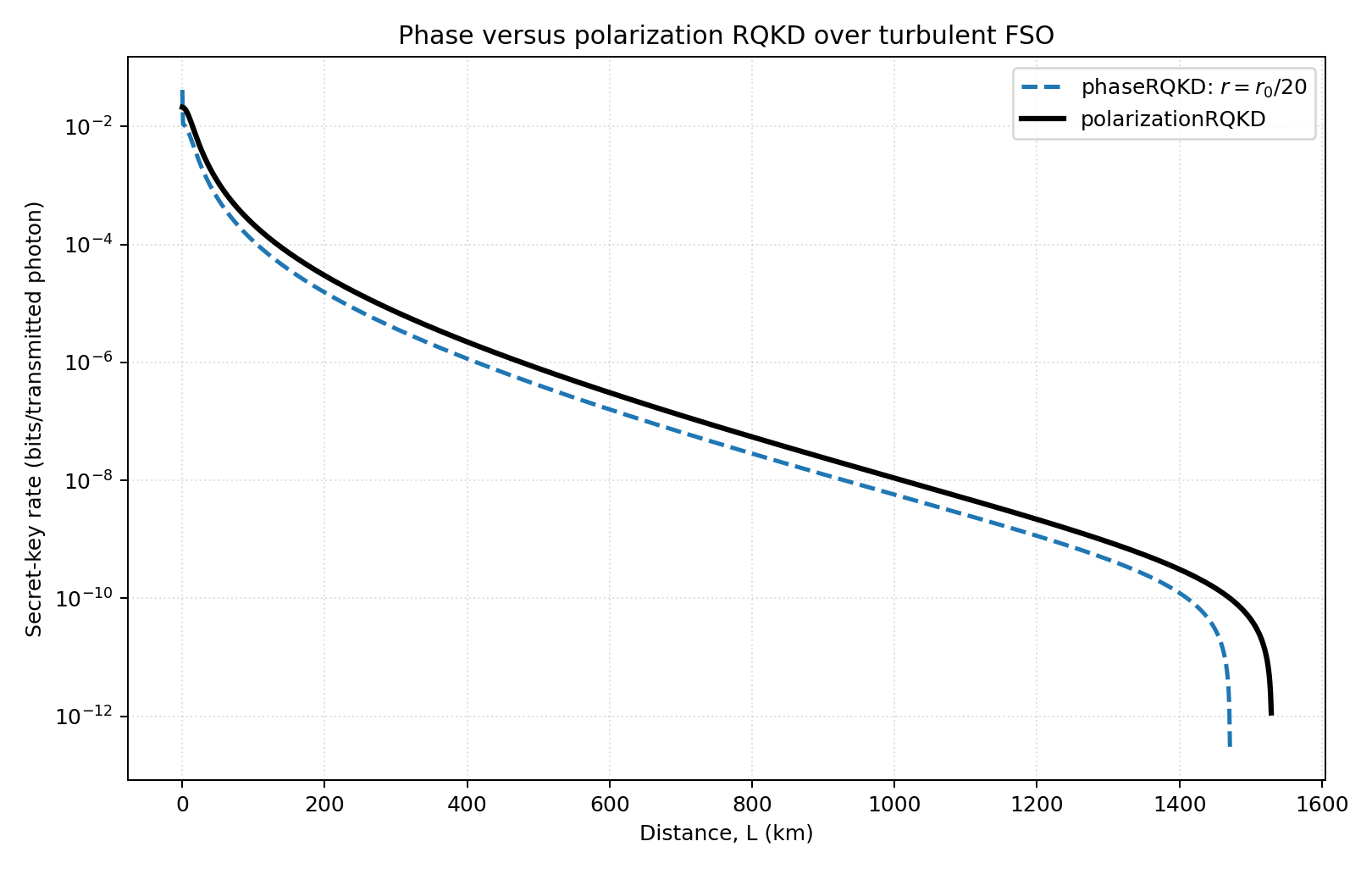}
	\caption{Secret-key rate versus transmission distance
		for the phase-encoded \cite{kravtsov,bebrov} and polarization-encoded RQKD
		protocols over a turbulent FSO channel.
		For the phase-encoded protocol, the phase-turbulence
		configuration $r=r_0/20$ is considered.
		The secret-key rate of the phase-encoded protocol
		is evaluated following Ref.~\cite{bebrov},
		taking into account the error-rate model presented
		in appendix~\ref{er-phase}.
		Polarization distortions induced by atmospheric
		turbulence are assumed to be negligible.}
	\label{fig:rate_comparison}
\end{figure*}

\indent Figure~\ref{fig:rate_comparison} presents the
secret-key rate of the proposed polarization-encoded
RQKD protocol and compares its performance with that
of the phase-encoded RQKD protocol. The secret-key rate
of the standard phase-encoded RQKD protocol is given by
\begin{equation}
	R_{\mathrm{ph}}
	=
	Q_{\mathrm{ph}}
	\left[1-\chi_{\mathrm{phase}}-fH_2(e_{\mathrm{phase}})\right],
	\label{eq:key-rate-phase}
\end{equation}
where $Q_{\mathrm{ph}}$ denotes the conclusive
single-photon yield, given by
\begin{equation}
	Q_{\mathrm{ph}}
	=
	Y_0^{\mathrm{ph}}
	+p_{\mathrm{det}}^{\mathrm{ph}}\eta,
\end{equation}
with $Y_0^{\mathrm{ph}}$ denoting the background yield
associated with the conclusive detectors and
$p_{\mathrm{det}}^{\mathrm{ph}}$ the probability of a
signal-induced conclusive event in the absence of channel
losses and detector inefficiency.
The corresponding QBER is denoted by $e_{\mathrm{phase}}$. The information leakage $\chi_{\mathrm{phase}}$ is given in Ref.~\cite{bebrov}.
A detailed mathematical description of the error-rate
model of the phase-encoded protocol is provided in
appendix~\ref{er-phase}.
To ensure a consistent comparison, both protocols are
evaluated under identical channel conditions and within
the same information-leakage model.\\
\indent As expected, the secret-key rates of both protocols
decrease with increasing transmission distance owing to
atmospheric attenuation, beam spreading, and the increasing
relative contribution of dark counts to the QBER.\\
\indent Unlike its polarization-encoded counterpart,
the phase-encoded RQKD protocol is inherently sensitive
to relative phase fluctuations induced by atmospheric
turbulence, which affect both the probability of a
conclusive detection event and the QBER~\cite{bebrov}.
Within the adopted detection model, the polarization-encoded
protocol exhibits no dependence on the phase-fluctuation
variance (see section \ref{math}). Consequently, despite its lower intrinsic
probability of a conclusive detection event, the
polarization-encoded protocol achieves a higher secret-key
rate over almost the entire transmission range considered.\\
\indent The advantage of polarization encoding becomes
particularly pronounced at large transmission distances,
where the increasing relative contribution of dark counts,
combined with turbulence-induced phase errors in the
phase-encoded protocol, causes its secret-key rate to
vanish at a shorter transmission distance.\\
\indent The maximum transmission distances, determined
by the condition of a positive secret-key rate, are
approximately 1470~km and 1530~km for the phase-encoded
and polarization-encoded RQKD protocols, respectively.
Thus, under the channel conditions considered,
polarization encoding extends the maximum transmission
distance by approximately 60~km, corresponding to
an improvement of about 4\%.\\
\indent These results demonstrate that, within the adopted
channel, error-rate, and information-leakage models,
polarization encoding enhances the robustness of RQKD
against atmospheric phase fluctuations. In particular,
even for the relatively small ratio $r/r_0=1/20$,
the proposed protocol achieves both a higher secret-key
rate over almost the entire transmission range and a
longer maximum transmission distance than its
phase-encoded counterpart.

\begin{table*}[!hbt]
	\caption{\label{table-parameters} Parameters used in evaluating the key rate of a single-photon (and coherent-state) RQKD over a turbulent FSO. Values and expressions below are taken in accordance with Refs. \cite{fried,dayton,anguita,peters,cardoso,parra,liao}. }
	\begin{tabular}{c|c}
		\hline
		\textbf{Quantity} & \textbf{Value}/\textbf{Expression}\\
		\hline\hline
		\textit{Wavenumber} & $k=\frac{2\pi}{\lambda}$; $\lambda$ measured in [m]\\
		\textit{Wavelength} & $\lambda=850\times 10^{-9}$[m] ($850$[nm])\\
		\textit{Rytov number} & $\sigma_{R}^2 = 1.23C^2_nk^{\frac{7}{6}}L^{\frac{11}{6}}$;  $L$ measured in [m]\\
		\textit{Stucture constant} & $C^2_n = 10^{-18}$ \\
		\textit{Transmissivity} & $\eta=\eta_{\text{det}}\eta_{\text{lt}}\eta_{\text{atm}}$\\
		\textit{Detector efficiency} & $\eta_{\text{det}}=0.5$\\
		\textit{Dark count rate} & $p_{\text{dark}} = 10^{-10}$\\
		\textit{Turbulent link transmissivity} & $\eta_{\text{lt}} = 1-\mathrm{e}^{-
			\frac{2a_R^{2}}{w_{\text{lt}}^2}}$; $a_R$ measured in [m]\\
		\textit{Atmospheric loss} & $\eta_{\text{atm}}=\mathrm{e}^{-L\alpha_0\mathrm{e}^{-\frac{h_0}{6600}}}$; $h_0$, $L$ measured in [m]\\
		\textit{Receiver aperture radius} & $a_R=0.05$[m] ($5$[cm])\\
		\textit{Long-term beam waist at the receiver} & $w_{\text{lt}}=w_L\sqrt{1+\frac{4}{3}q\times\Lambda}$\\
		\textit{Beam's spot size} & $w_L=\sqrt{w_0^2\left[\left(1-\frac{L}{R_0}\right)^2+\left(\frac{L}{L_R}\right)^2\right]}$; $L$ measured in [m]\\
		\textit{Beam's Rayleigh length} & $L_R =\frac{\pi w_0^2}{\lambda}$; $w_0$ measured in [m]\\
		\textit{Initial spot size} (\textit{radius}) & $w_0 = 0.05$[m] ($5$[cm])\\
		\textit{Initial spot curvature} & $R_0 = \infty$ (collimated beam)\\
		\textit{Gaussian beam parameter} & $\Lambda = \frac{2L}{kw_L^2}$; $L$ measured in [m]\\
		\textit{$q$ parameter} & $q = 0.74\sigma_R^2Q_m^{\frac{1}{6}}$, $Q_m=35.05\frac{L}{k\times l_0^2}$; $L$, $l_0$ measured in [m]\\
		\textit{Extinction factor at sea level} & $\alpha_0 = 5\times 10^{-6}$[m$^{-1}$] \\
		\textit{Altitude} & $h_0 = 30$[m]\\
		\textit{Turbulence inner scale} & $l_0 = 0.001$[m] ($1$[mm])\\
		\textit{Phase structure function} & $D_\phi = 6.88\left(\frac{r}{r_0}\right)^{\frac{5}{3}}=\sigma^2_\phi$\\
		\textit{Fried's turbulent parameter}& $r_0=\left(0.423C^2_nk^2L\right)^{-\frac{3}{5}}$; $L$ measured in [m] \\
		\textit{Interferometer arms separation} &  $r$ ($r>r_0$ \textit{or} $r=r_0$ \textit{or} $r<r_0$)
	\end{tabular}
\end{table*}

\section{Summary}
\label{summary}
\indent In this work, we have proposed a polarization-encoded
relativistic quantum key distribution protocol aimed at
improving the robustness of RQKD against atmospheric
phase fluctuations.
The proposed protocol preserves the underlying communication
mechanism of standard RQKD, namely the space-time spreading
and subsequent localization of a single-photon quantum
state, while introducing an alternative encoding and
measurement strategy based on two nonorthogonal
polarization states.\\
\indent The distinctive feature of the proposed protocol
consists of adapting the polarization measurement basis
to the polarization selected by the receiving party.
The mathematical analysis of the detection statistics
has shown that, under ideal polarization preparation
and analysis, the probability of a conclusive detection
event is independent of the relative optical phase.
Consequently, atmospheric phase fluctuations do not
contribute to the quantum bit error rate within the
adopted phase-disturbance model.
This property is of particular importance for free-space
optical communication, where maintaining stable
interference conditions may be challenging.\\
\indent The proposed protocol has also been examined from
the standpoint of relativistic security.
By analyzing the spacetime intervals associated with
the propagation of the quantum system, we have demonstrated
that the considered intercept-and-resend strategy,
which seeks to preserve the intended operation of the
receiver, is incompatible with relativistic causality.
A causally admissible modification of this strategy,
on the other hand, disrupts the mechanism underlying
the establishment of correlated detection outcomes.\\
\indent Particular attention has been devoted to the
influence of practical implementation imperfections.
An analytical error-rate model accounting for imperfect
polarization discrimination and background-induced
detection events has been developed.
To mitigate polarization leakage, a cascaded polarization
analyzer (PBS followed by a polarizer) has been considered.
Under appropriate alignment and polarization-preserving
leakage conditions, the contribution of imperfect
polarization discrimination to the QBER can be
substantially reduced, albeit at the expense of
additional optical insertion losses.\\
\indent The performance of the proposed protocol has been
evaluated in terms of the achievable secret-key rate
over a turbulent free-space optical channel.
Despite its lower intrinsic probability of a conclusive
detection event, the polarization-encoded protocol
achieves a higher secret-key rate than its phase-encoded
counterpart over almost the entire transmission range
considered.
For the channel and system parameters employed in the
numerical analysis, the maximum transmission distance
is extended from approximately 1470~km to 1530~km,
corresponding to an improvement of about 4\%.\\
\indent Within the adopted channel, error-rate, and
information-leakage models, the results indicate that
polarization encoding represents a promising alternative
to phase encoding in relativistic quantum key distribution,
particularly for applications involving long-distance
free-space optical communication.
Further investigations may address the influence of
atmospheric polarization distortions, the experimental
realization of the proposed receiver, and the extension
of the security analysis to more general attack strategies.


\bibliography{sn-bibliography}

@INPROCEEDINGS{Bennett1984,
	author    = " Bennett, C. and  Brassard, G.",
	title     = "Quantum cryptography: Public key distribution and coin tossing",
	booktitle = "International Conference on Computers, Systems \& Signal Processing",
	volume    = "1",
	pages     = "175-179",
	year      = "1984",
}

@ARTICLE{Lo2012,
	author       = "Lo,H.-K.  and  Curty, M. and Qi,B. ", 
	title        = "Measurement-Device-Independent Quantum Key Distribution",
	year         = "2012", 
	journal      = "Phys.\ Rev. Lett.", 
	volume       = "108", 
	pages        = "130503",
}

@ARTICLE{parra,
	author       = "Jordan-Parra, J. and Garcia-Romero, M. Gil-Lopez, J. and Ruiz-De-Azua, J. and Paradells, J.",
	title		 = "Satellite Quantum Key Distribution: Analysis, Modeling, Performance Evaluation, and Validation",
	journal      = "IEEE Open Journal of the Communications Society",
	volume       = "6",
	pages        = "8241-8271",
	year         = "2025",
	doi 		 = "10.1109/OJCOMS.2025.3614698"
}

@ARTICLE{liao,
	author       = "Liao, Sh.-K. and Cai, W.-Q. and Liu, W.-Y. and Zhang, L. and Li, Y. and Ren, J.-G. and Yin, J. and Shen, Q. and Cao, Y. and Li, Zh.-P. and Li, F.-Zh. and Chen, X.-W. and Sun, L.-H. and Jia, J.-J. and Wu, J.-C. and Jiang, X.-J. and Wang, J.-F. and Huang, Y.-M. and Wang, Q. and Zhou, Y.-L. and Deng, L. and Xi, T. and Ma, L. and Hu, T. and Zhang, Q. and Chen, Y.-A. and Liu, N.-L. and Wang, X.-B. and Zhu, Zh.-C. and Lu, C.-Y. and Shu, R. and Peng, C.-Z. and Wang, J.-Y. and Pan,  J.-W.",
	title		 = "Satellite-to-ground quantum key
	distribution",
	journal      = "Nature",
	volume       = "549",
	pages        = "43-59",
	year         = "2017",
	doi			 = "10.1038/nature23655"
}

@BOOK{nielsen,
	author       = {M. Nielsen and I. Chuang},
	year         = 2010,
	title        = {Quantum Computation and Quantum Information},
	publisher    = {Cambridge University Press},
	doi		= "10.1017/CBO9780511976667"
}

@article{minkowski1915,
	author  = {Minkowski, Hermann},
	title   = {Das Relativit{\"a}tsprinzip},
	journal = {Annalen der Physik},
	volume  = {352},
	number  = {15},
	pages   = {927--938},
	year    = {1915}
}

@book{taylor1992,
	author    = {Taylor, E. and Wheeler, J.},
	title     = {Spacetime Physics: Introduction to Special Relativity},
	edition   = {2},
	publisher = {W. H. Freeman},
	address   = {New York},
	year      = {1992}
}

@ARTICLE{b92,
	author       = "Ch. Bennett",
	year         = "1992",
	journal      = "Phys.\ Rev. Lett.",
	volume       = "68",
	pages        = "3121",
	doi		= "10.1103/PhysRevLett.68.3121"
}

@ARTICLE{ekert,
	author       = "A. Ekert",
	year         = "1991",
	journal      = "Phys.\ Rev. Lett.",
	volume       = "67",
	pages        = "661",
	doi		= "10.1103/PhysRevLett.67.661"
}

@ARTICLE{hwang,
	author       = "W.-Y. Hwang",
	year         = "2003",
	journal      = "Phys.\ Rev. Lett.",
	volume       = "91",
	pages        = "057901",
	doi		= "10.1103/PhysRevLett.91.057901"
}

@ARTICLE{lima,
	author       = "Lima, D. and Rigolin, G.",
	title		 = "Asymptotic security analysis of teleportation-based quantum cryptography",
	year         = "2020",
	journal      = "Quant. Inf. Process.",
	volume       = "19",
	pages        = "201",
	doi		= "10.1007/s11128-020-02701-w"
}

@ARTICLE{lucamarini,
	author       = "M. Lucamarini and Z. Yuan and J. Dynes and A. Shields",
	year         = "2018",
	journal      = "Nature",
	volume       = "557",
	pages        = "400",
	doi		= "10.1038/s41586-018-0066-6"
}

@ARTICLE{sns,
	author       = "X.-B. Wang and Z.-W. Yu and X.-L. Hu",
	year         = "2018",
	journal      = "Phys.\ Rev. A",
	volume       = "98",
	pages        = "062323",
	doi		= "10.1103/PhysRevA.98.062323"
}

@ARTICLE{kravtsov,
	author       = "Kravstov, K. and Radchenko, I. and Kulik, S. and Molotkov, S.",
	title		 = "Relativistic quantum key
	distribution system with one-way quantum communication",
	year         = "2018",
	journal      = "Scientific Reports",
	volume       = "8",
	pages        = "6102",
	doi		= "10.1038/s41598-018-24533-6"
}

@ARTICLE{kulik,
	author       = "Kulik, S. and Molotkov, S.",
	title		 = "Relativistic quantum key
	distribution between stationary objects based on coherent states",
	year         = "2026",
	journal      = "Uspekhi Fizicheskikh Nauk",
	volume       = "196",
	pages        = "1086-1097",
	doi		= "10.3367/UFNr.2026.08.040188"
}

@ARTICLE{molotkov,
	author       = "Molotkov, S.",
	title		 = "Relativistic quantum cryptography",
	year         = "2011",
	journal      = "J. Exp. Theor. Phys.",
	volume       = "112	",
	pages        = "370-379",
	doi		= "10.1134/S106377611102018X"
}

@ARTICLE{toyoshima,
	author       = "Toyoshima, M. and Takenaka, H. and Shoji, Y. and Takayama, Y. and Koyama, Y. and Kunimori, H.",
	title		 = "Polarization measurements through space-to-ground atmospheric propagation paths by using a highly polarized laser source in space",
	year         = "2009",
	journal      = "Optics Express",
	volume       = "17",
	pages        = "22333-22340",
	doi		= "10.1364/OE.17.022333"
}

@ARTICLE{bebrov2,
	author       = "Bebrov, G.",
	title		 = "Relativistic quantum key
	distribution in the single-photon regime",
	year         = "2025",
	journal      = "Geom. Integrability \& Quantization",
	volume       = "32",
	pages        = "1",
	doi			 = "10.7546/giq-32-2025-1-17"
}

@MISC{bebrov,
	author       = "Bebrov, G.",
	title		 = "Key Evaluation of a Relativistic Quantum Key Distribution Over a Turbulent Free-space 	Optical Channel",
	archivePrefix = {arXiv},
	primaryClass  = {quant-ph},	
	doi			 = "arXiv:2610.38808"
}

@ARTICLE{fried,
	author       = "Fried, D.",
	title		 = "Optical Resolution Through a Randomly Inhomogeneous Medium for Very Long and Very Short Exposures",
	year         = "1966",
	journal      = "Journal of the Optical Society of America",
	volume       = "56",
	pages        = "1372",
	doi			 = "10.1364/JOSA.56.001372"
}

@ARTICLE{dayton,
	author       = "Dayton, D. and Pierson, B. and Spielbusch, B. and Gonglewski, J.",
	title		 = "Atmospheric structure function measurements with a Shack-Hartmann wave-front sensor",
	year         = "1992",
	journal      = "Optics Letters",
	volume       = "17",
	pages        = "1737",
	doi			 = "10.1364/OL.17.001737"
}

@ARTICLE{anguita,
	author       = "Anguita, J. and Neifeld, M. and Vasic, B.",
	title		 = "Turbulence-induced channel crosstalk in an
	orbital angular momentum-multiplexed free-space optical link",
	year         = "2008",
	journal      = "Applied Optics",
	volume       = "47",
	pages        = "2414",
	doi			 = "10.1364/AO.47.002414"
}

@ARTICLE{peters,
	author       = "Peters, C. and Cocotos, V. and Forbes, A.",
	title		 = "Structured light in atmospheric
	turbulence—a guide to its digital implementation: tutorial",
	year         = "2025",
	journal      = "Advances in Optics and Photonics",
	volume       = "17",
	pages        = "113",
	doi 		 = "10.1364/AOP.538883"
}

@ARTICLE{cardoso,
	author       = "Aguilar-Cardoso, A. and Li, Ch. and Luck, T. and Ferrer-Garcia, M. and Upham, J. and Lundeen, J. and Boyd, R.",
	title		 = "All-optical turbulence mitigation for free-space quantum key distribution using stimulated
	parametric down-conversion",
	year         = "2026",
	journal      = "Optica",
	volume       = "13",
	pages        = "386",
	doi 		 = "10.1364/OPTICA.583778"
}

@ARTICLE{wang2005,
	author       = "Wang, X.-B.",
	journal      = "Phys. Rev. Lett.",
	volume       = "94",
	pages        = "230503",
	year         = "2005",
}

@ARTICLE{wang2,
	author       = "Wang, X.-B.",
	journal      = "Phys. Rev. A",
	volume       = "72",
	pages        = "012322",
	year         = "2005",
}

@ARTICLE{ma2005,
	author       = "Ma, X. and Qi, B. and Zhao, Y. and Lo, H.-K.",
	title		 = "Practical decoy state for quantum key distribution",
	journal      = "Phys. Rev. A",
	volume       = "72",
	pages        = "012326",
	year         = "2005",
	doi			 = "10.1103/PhysRevA.72.012326"
}

@ARTICLE{lin-lutken,
	author       = "Lin, J and L{\"u}tkenhaus, N.",
	title		 = "Simple security analysis of phase-matching measurement-device-independent quantum key distribution",
	journal      = "Phys. Rev. A",
	volume       = "98",
	pages        = "042332",
	year         = "2018",
	doi			 = "10.1103/PhysRevA.98.042332"
}

\appendix
\section{Mathematical description}
\label{math}
\indent Initially \textit{Alice} prepares a horizontally polarized single photon
\begin{equation*}
	\begin{split}
		\ket{\psi_{\text{in}}} = a^\dagger_{\text{H}}\ket{0}
	\end{split}
\end{equation*}
The photon is incident on a balanced beam splitter BS$_s$, which produces a superposition of the upper- and lower-arm spatial modes
\begin{equation*}
	\begin{split}
		a^\dagger_{\text{H}}\ket{0}\xrightarrow{\text{BS}_s} \frac{1}{\sqrt{2}}(u^\dagger_{\text{H}} + l^\dagger_{\text{H}})\ket{0} = \ket{\psi_s}.
	\end{split}
\end{equation*}
Here $u^\dagger_{\text{\sffamily p}}$ and $l^\dagger_{\text{\sffamily p}}$ denote single-photon creation operators associated with the upper and lower arms, respectively, with polarization $\text{\sffamily p}$.\\
\indent With probability $p_a = 1/2$ \textit{Alice} randomly applies a polarization Hadamard operator $\hat{H}_a$ to the upper-arm mode. Recall that the Hadamard transformation is defined by
\begin{equation*}
	\begin{split}
		\hat{H}\ket{H} = \ket{D} \quad \hat{H}\ket{V} = \ket{A}
	\end{split}
\end{equation*}
where 
\begin{equation*}
	\begin{split}
		\ket{D} = \frac{\ket{H}+\ket{V}}{\sqrt{2}}, \quad \ket{A} = \frac{\ket{H}-\ket{V}}{\sqrt{2}}
	\end{split}
\end{equation*}
The resulting quantum state is described by the ensemble
\begin{equation*}
	\begin{split}
		\sigma_a = \frac{1}{2}(\ketbra{\psi_0} + \ketbra{\psi_1})
	\end{split}
\end{equation*}
where
\begin{equation*}
	\begin{split}
		\ket{\psi_0} &= \frac{1}{\sqrt{2}}(u^\dagger_{\text{H}} + l^\dagger_{\text{H}})\ket{0},\\
		\ket{\psi_1} &= \frac{1}{\sqrt{2}}(u^\dagger_{\text{D}}+l^\dagger_{\text{H}})\ket{0}.
	\end{split}
\end{equation*}
The first state corresponds to \textit{Alice} not applying $\hat{H}_a$, whereas the second corresponds to applying it. Notice that both conditional states are pure (take into account appendix \ref{pure}), although their statistical mixture $\sigma_a$ is mixed. \\
\indent The upper- and lower-arm modes independently acquire phase shifts $\phi_u$ and $\phi_l$, respectively. Taking into account mirror reflections, the resulting states can be written as
\begin{equation*}
	\begin{split}
		\ket{\psi'_0} &= \frac{1}{\sqrt{2}}(\mathrm{e}^{\mathrm{i}\phi_u}u^\dagger_{\text{H}}+\mathrm{e}^{\mathrm{i}\phi_l}l^\dagger_{\text{H}}),\\
		\ket{\psi'_1} &= \frac{1}{\sqrt{2}}(\mathrm{e}^{\mathrm{i}\phi_u}u^\dagger_{\text{D}}+\mathrm{e}^{\mathrm{i}\phi_l}l^\dagger_{\text{H}}).
	\end{split}
\end{equation*}
Here $\phi_u$ and $\phi_l$ are understood to include any differential phase contributions from propagation and reflection. A common global phase is omitted. The density operator becomes
\begin{equation*}
	\begin{split}
		\sigma'_a = \frac{1}{2}(\ketbra{\psi'_0} + \ketbra{\psi'_1}).
	\end{split}
\end{equation*}
\indent Independently of \textit{Alice}'s choice, \textit{Bob} randomly applies a polarization Hadamard operator $\hat{H}_b$ to the lower-arm mode. Consequently the quantum system is described by four possible states
\begin{equation*}
	\begin{split}
		\sigma_b = \frac{1}{4}\sum\limits_{a,b\in\{0,1\}} \ketbra{\psi_{ab}}.
	\end{split}
\end{equation*}
Explicitly
\begin{equation*}
	\begin{split}
		\ket{\psi_{00}} &= \frac{1}{\sqrt{2}}(\mathrm{e}^{\mathrm{i}\phi_u}u^\dagger_{\text{H}}+\mathrm{e}^{\mathrm{i}\phi_l}l^\dagger_{\text{H}}),\\
		\ket{\psi_{01}} &= \frac{1}{\sqrt{2}}(\mathrm{e}^{\mathrm{i}\phi_u}u^\dagger_{\text{H}}+\mathrm{e}^{\mathrm{i}\phi_l}l^\dagger_{\text{D}}),\\
		\ket{\psi_{10}} &= \frac{1}{\sqrt{2}}(\mathrm{e}^{\mathrm{i}\phi_u}u^\dagger_{\text{D}}+\mathrm{e}^{\mathrm{i}\phi_l}l^\dagger_{\text{H}}),\\
		\ket{\psi_{11}} &= \frac{1}{\sqrt{2}}(\mathrm{e}^{\mathrm{i}\phi_u}u^\dagger_{\text{D}}+\mathrm{e}^{\mathrm{i}\phi_l}l^\dagger_{\text{D}}).  
	\end{split}
\end{equation*}
The first subscript identifies \textit{Alice}'s Hadamard choice, while the second identifies \textit{Bob}'s. Defining the relative phase
\begin{equation*}
	\begin{split}
		\Delta\phi = \phi_l - \phi_u
	\end{split}
\end{equation*}
and omitting the global factor $\mathrm{e}^{\mathrm{i}\phi_u}$, the above states can be expressed compactly as
\begin{equation*}
	\begin{split}
		\ket{\psi_{ab}} = \frac{1}{\sqrt{2}}(u^\dagger_{\text{\sffamily p}_a} +\mathrm{e}^{\mathrm{i}\Delta\phi}l^\dagger_{\text{\sffamily p}_b})\ket{0}
	\end{split}
\end{equation*}
where $\text{\sffamily p}_0$ $=$ $H$ and $\text{\sffamily p}_1$ $=$ $D$. \\
\indent The upper- and lower-arm modes interfere at the balanced beam splitter BS$_f$. The modes transform as follows
\begin{equation*}
	\begin{split}
		u^\dagger_{\text{\sffamily p}} &\rightarrow \frac{1}{\sqrt{2}}(c^\dagger_{\text{\sffamily p}}+d^\dagger_{\text{\sffamily p}}),\\
		l^\dagger_{\text{\sffamily p}} &\rightarrow \frac{1}{\sqrt{2}}(c^\dagger_{\text{\sffamily p}}-d^\dagger_{\text{\sffamily p}})
	\end{split}
\end{equation*}
where $c^\dagger$ and $d^\dagger$ correspond to upper and lower outputs (of BS$_f$), respectively. Therefore
\begin{equation*}
	\begin{split}
		\ket{\tilde{\psi}_{ab}}=\frac{1}{2}[c^\dagger_{\text{\sffamily p}_a} + \mathrm{e}^{\mathrm{i}\Delta\phi}c^\dagger_{\text{\sffamily p}_b}+d^\dagger_{\text{\sffamily p}_a}-\mathrm{e}^{\mathrm{i}\Delta\phi}d^\dagger_{\text{\sffamily p}_b}]\ket{0}.
	\end{split}
\end{equation*}
The four outputs state are then
\begin{equation*}
	\begin{split}
		\ket{\tilde{\psi}_{00}} &= \frac{1}{2}[(1+\mathrm{e}^{\mathrm{i}\Delta\phi})c^\dagger_{\text{H}} + (1-\mathrm{e}^{\mathrm{i}\Delta\phi})d^\dagger_{\text{H}}]\ket{0},\\
		\ket{\tilde{\psi}_{01}} &= \frac{1}{2}[c^\dagger_{\text{H}} + \mathrm{e}^{\mathrm{i}\Delta\phi}c^\dagger_{\text{D}} + d^\dagger_{\text{H}} - \mathrm{e}^{\mathrm{i}\Delta\phi}d^\dagger_{\text{D}}]\ket{0},\\
		\ket{\tilde{\psi}_{10}} &= \frac{1}{2}[c^\dagger_{\text{D}} + \mathrm{e}^{\mathrm{i}\Delta\phi}c^\dagger_{\text{H}} + d^\dagger_{\text{D}} - \mathrm{e}^{\mathrm{i}\Delta\phi}d^\dagger_{\text{H}}]\ket{0},\\
		\ket{\tilde{\psi}_{11}} &= \frac{1}{2}[(1+\mathrm{e}^{\mathrm{i}\Delta\phi})c^\dagger_{\text{D}} + (1-\mathrm{e}^{\mathrm{i}\Delta\phi})d^\dagger_{\text{D}}]\ket{0}.
	\end{split}
\end{equation*}
The output density operator has the form
\begin{equation*}
	\begin{split}
		\rho_{\text{out}} = \frac{1}{4}\sum\limits_{a,b\in\{0,1\}}\ketbra{\tilde{\psi}_{ab}}.
	\end{split}
\end{equation*}
\indent Unlike the standard RQKD (phase-encoded scheme), the proposed configuration employs a polarization beam splitter at each output of BS$_f$. \textit{Bob} configures the PBSs according to \textit{his} Hadamard choice: if $b=0$, both PBSs measure in the $\{H,V\}$ basis; if $b=1$, both PBSs measure in the $\{D,A\}$ basis. The corresponding projectors are
\begin{equation*}
	\begin{split}
		\Pi^{\text{\sffamily u}}_{\text{\sffamily p}} = c^\dagger_{\text{\sffamily p}}\ketbra{0}c_{\text{\sffamily p}}, \quad \Pi^{\text{\sffamily l}}_{\text{\sffamily p}} = d^\dagger_{\text{\sffamily p}}\ketbra{0}d_{\text{\sffamily p}}.
	\end{split}
\end{equation*}
The probability of a detector click, conditioned on the configuration ($a,b$), is
\begin{equation*}
	\begin{split}
		P(\text{\sffamily p}^j|a,b) = \Tr(\Pi^{j}_{\text{\sffamily p}}\ketbra{\tilde{\psi}_{ab}})
	\end{split}
\end{equation*} 
where $j \in \{\text{\sffamily u},\text{\sffamily l}\}$. \\
\indent Denote the key-generating orthogonal-polarization outputs (orthogonal to the transferred ones) by
\begin{equation*}
	\begin{split}
		\text{\sffamily p}_{\perp} = \begin{cases}
			V, & b=0,\\
			A, & b=1.
		\end{cases}
	\end{split}
\end{equation*}
The corresponding projectors are $\Pi^{\text{\sffamily u}}_{\text{\sffamily p}_{\perp}}$ and $\Pi^{\text{\sffamily l}}_{\text{\sffamily p}_\perp}$. \\
\indent Let 
\begin{equation*}
	\begin{split}
		s_{ab} = \braket{\text{\sffamily p}_a}{\text{\sffamily p}_b} = \begin{cases}
			1, & a=b,\\
			1/\sqrt{2}, & a \neq b.
		\end{cases}
	\end{split}
\end{equation*}
Then the conditional detection probabilities are
\begin{equation*}
	\begin{split}
		P(\text{\sffamily p}^{\text{\sffamily u}}_b|a,b) &= \frac{1}{4}(1 + s^2_{ab} + 2s_{ab}\cos\Delta\phi),\\
		P(\text{\sffamily p}^{\text{\sffamily l}}_b|a,b) &= \frac{1}{4}(1 + s^2_{ab} - 2s_{ab}\cos\Delta\phi)
	\end{split}
\end{equation*}
and 
\begin{equation*}
	\begin{split}
		P(\text{\sffamily p}^{\text{\sffamily u}}_\perp|a,b) = P(\text{\sffamily p}^{\text{\sffamily l}}_\perp|a,b) = \frac{1}{4}(1 - s^2_{ab}).
	\end{split}
\end{equation*}
For $\Delta\phi = 0$, the results are those presented in table \ref{results}. Note that the probabilities in table \ref{results} are conditional on the corresponding configuration (modes). Take into account that each configuration occurs with probability $1/4$ ($p_a\cdot p_b$ given $p_a = p_b = 1/2$). Notice that the results presented in table \ref{probabilities} are obtained by multiplying the conditional probabilities by a factor of $1/4$.
\begin{table}[!hbt] 
	\caption{\label{results} Conditional detection probabilities given upper- and lower-arm modes at BS$_f$ (configurations). Polarizations $H$ and $D$ are referred to \textit{parallel} ($\parallel$), whereas $V$ and $A$ are referred to \textit{orthogonal} ($\perp$). Recall that $\{H,V\}$ measurement basis is used if $l^\dagger_{\text{H}}$, and  $\{D,A\}$ measurement basis is used if $l^\dagger_{\text{D}}$.}
	\begin{ruledtabular}
		\begin{tabular}{ccccc}
			\textbf{modes}&\textbf{upper} $\parallel$&\textbf{upper} $\perp$ &\textbf{lower} $\parallel$ & \textbf{lower} $\perp$ \\
			\hline
			($u^\dagger_{\text{H}}$,$l^\dagger_{\text{H}}$)& $100\%$ & $0\%$ & $0\%$ & $0\%$\\
			($u^\dagger_{\text{H}}$,$l^\dagger_{\text{D}}$)& $72.86\%$ & $12.5\%$ & $2.14\%$ & $12.5\%$ \\
			($u^\dagger_{\text{D}}$,$l^\dagger_{\text{H}}$)& $72.86\%$ & $12.5\%$ & $2.14\%$ & $12.5\%$\\
			($u^\dagger_{\text{D}}$,$l^\dagger_{\text{D}}$)& $100\%$ & $0\%$ & $0\%$ & $0\%$ \\
		\end{tabular}
	\end{ruledtabular}
\end{table}
\\ \indent As introduced in section \ref{protocol}, \textit{Bob} retains clicks at the polarization output orthogonal to \textit{his} selected measurement basis. So the key-event projector for a fixed choice $b$ is
\begin{equation*}
	\begin{split}
		\Pi^{(b)}_{\text{key}} = \Pi^{\text{\sffamily u}}_{\text{\sffamily p}_\perp} + \Pi^{\text{\sffamily l}}_{\text{\sffamily p}_\perp}.
	\end{split}
\end{equation*}
The overall probability of obtaining a key-generating detection event is then
\begin{equation*}
	\begin{split}
		p_{\text{det}} &= \frac{1}{4}\sum\limits_{a,b} \Tr(\Pi^{(b)}_{\text{key}}\ketbra{\tilde{\psi}_{ab}})\\
		&= \frac{1}{4}(0 + \frac{1}{4} + \frac{1}{4}) = \frac{1}{8}. 
	\end{split}
\end{equation*}
It is really worth noting that the orthogonal-polarization detection probability is independent of the relative phase $\Delta\phi$.\\
\indent We now analyze the behavior of the proposed scheme in the case of turbulent FSO channel. We assume \cite{bebrov}
\begin{equation*}
	\begin{split}
		\phi_u &\sim \mathcal{N}(0,\sigma^2_\phi),\\
		\phi_l &\sim \mathcal{N}(0,\sigma^2_\phi)
	\end{split}
\end{equation*}
with independent fluctuations. Consequently
\begin{equation*}
	\begin{split}
		\Delta\phi = \phi_l - \phi_u \sim \mathcal{N}(0,2\sigma^2_\phi).
	\end{split}
\end{equation*}
This yields \cite{bebrov}
\begin{equation*}
	\begin{split}
		\langle \cos\Delta\phi \rangle = \mathrm{e}^{-\sigma^2_\phi}.
	\end{split}
\end{equation*}
Thus the phase averaging modifies the parallel-polarization detection probabilities:
\begin{equation*}
	\begin{split}
		\langle P(\text{\sffamily p}^{\text{\sffamily u}}_b|a,b) \rangle &= \frac{1}{4}(1 + s^2_{ab} + 2s_{ab}\mathrm{e}^{-\sigma^2_\phi}),\\
		\langle P(\text{\sffamily p}^{\text{\sffamily l}}_b|a,b) \rangle &= \frac{1}{4}(1 + s^2_{ab} - 2s_{ab}\mathrm{e}^{-\sigma^2_\phi}).
	\end{split}
\end{equation*}
However the orthogonal-polarization probabilities remain unchanged:
\begin{equation*}
	\begin{split}
		\langle P(\text{\sffamily p}^{j}_\perp|a,b) \rangle = \frac{1}{4}(1 - s^2_{ab}).
	\end{split}
\end{equation*}
Consequently
\begin{equation*}
	\begin{split}
		\langle p_{\text{det}} \rangle = \frac{1}{8}.
	\end{split}
\end{equation*}
as shown previously. This phase independence is a significant difference from the phase-encoded scheme, whose detection probability depends explicitly on the phase difference between the two arms (see \cite{bebrov} for reference).\\
\indent Under the ideal assumptions (perfect detectors, BSs, PBSs) considered here, the configurations ($u^\dagger_{\text{H}}$,$l^\dagger_{\text{H}}$) $=$ $HH$ and ($u^\dagger_{\text{D}}$,$l^\dagger_{\text{D}}$) $=$ $DD$ do not produce orthogonal-polarization detections. Hence
\begin{equation*}
	\begin{split}
		\langle P(\text{\sffamily p}^{j}_\perp|HH) =  P(\text{\sffamily p}^{j}_\perp|DD) = 0.
	\end{split}
\end{equation*}
If an erroneous key event is defined as an orthogonal-polarization click when \textit{Alice} and \textit{Bob} make identical Hadamard choices ($a=b$), then 
\begin{equation*}
	\begin{split}
		p_{\text{error}} = 0.
	\end{split}
\end{equation*}
Moreover, because phase fluctuations do not change the orthogonal-polarization probabilities, 
\begin{equation*}
	\begin{split}
		\langle p_{\text{error}} \rangle = 0
	\end{split}
\end{equation*}
even in the presence of a phase turbulence. This conclusion holds for pure scalar phase fluctuations, ideal polarization transformations, and ideal detection. Polarization drift, imperfect Hadamard operations, detector dark counts, and other optical imperfections can introduce non-zero error probabilities.

\section{Purity of $\rho_0$ and $\rho_{1}$}
\label{pure}
The \textit{purity} of a quantum state $\rho=\ketbra{\rho}$ is given by \cite{nielsen}
\begin{equation}\label{purity}
	\begin{split}
		\Gamma = \Tr(\rho^2).
	\end{split}
\end{equation}
This expression holds for \textit{normalized} states ($\braket{\rho}{\rho}=1$). Note that a state is pure \textit{if and only if} $\Gamma=1$. We first verify that $\ket{H}$ and $\ket{D}$ are normalized. Then we check if these states are \textit{pure} according to equation \eqref{purity}.\\
\indent Since
\begin{equation}\label{commutators}
	\begin{split}
		\commutator{u}{u^\dagger} &= 1,\\
		\commutator{l}{l^\dagger} &= 1,\\
		\commutator{u}{l^\dagger} &= 0,\\
	\end{split}
\end{equation}
we get
\begin{equation}
	\begin{split}
		\braket{H}{H} &= \frac{1}{2}\bra{0}(u_{\text{H}}+l_{\text{H}})(u^\dagger_{\text{H}}+l^\dagger_{\text{H}})\ket{0}=1
	\end{split}
\end{equation}
and 
\begin{equation}
	\begin{split}
		\braket{D}{D} &= \frac{1}{2}\bra{0}(u_{\text{D}}+l_{\text{D}})(u^\dagger_{\text{D}}+l^\dagger_{\text{D}})\ket{0}=1.
	\end{split}
\end{equation}
Hence both $\ket{H}$ and $\ket{D}$ are normalized. We find that 
\begin{equation}
	\begin{split}
		\rho_0^2 &= \ket{H}\braket{H}{H}\bra{H} = \ketbra{H} = \rho_0,\\
		\rho_{1}^2 &= \ket{D}\braket{D}{D}\bra{D} = \ketbra{D} = \rho_{1}
	\end{split}
\end{equation}
The last results imply
\begin{equation}
	\begin{split}
		\Gamma_0 &= \Tr(\rho_0^2) = \Tr(\rho_0) = 1,\\
		\Gamma_{1} &= \Tr(\rho_{1}^2) = \Tr(\rho_{1}) = 1.\\
	\end{split}
\end{equation}
Therefore $\rho_0$ and $\rho_{1}$ are pure.

\section{Error rate in polarizationRQKD}
\label{er-pol}
Following the approach of Ma \textit{et al.}~\cite{ma2005},
we distinguish between two contributions to the error rate:
background-induced errors and errors arising from imperfections
of the optical system. Since atmospheric turbulence induces
only minor polarization distortions~\cite{toyoshima}, the
corresponding contribution to the error rate is neglected throughout this work. \\
\indent Let $\epsilon_1$ denote the probability that a photon incident
on a polarization beam splitter (PBS) is routed to the
undesired output port (leakage probability), and let $t_1$ denote the transmission
probability associated with the desired output port.
Imperfect polarization discrimination may thus give rise
to conclusive detection events even when Alice and Bob
prepare identical polarization states (configurations that ideally yield no conclusive detection events).\\
\indent Taking into account the equal a priori probabilities of
the four possible polarization configurations ($HH$, $DD$, $HD$, and $DH$), the
probability of an erroneous conclusive event is given by
\begin{equation}
	p_{\mathrm{err}}^{\mathrm{pol}}
	= \frac{\epsilon_1}{2},
\end{equation}
whereas the overall probability of a conclusive event
amounts to
\begin{equation}
	p_{\mathrm{det}}^{\mathrm{pol}}
	= \frac{t_1+7\epsilon_1}{8}.
\end{equation}
The derivation of the last two quantities is given in appendix \ref{det-prob-pol}. \\
\indent The contribution of imperfect polarization discrimination
can be substantially reduced by employing a cascaded
polarization analyzer (as illustrated in figure \ref{setup}), wherein an additional polarizer
is placed at each conclusive output port of the PBS.
Let $\epsilon_2$ and $t_2$ denote the leakage and transmission
probabilities of the additional polarizer, respectively.
Provided that the polarization axes of the two optical
elements are properly aligned and that the leakage field
retains its polarization, the effective leakage and
transmission probabilities are given by
\begin{equation}
	\epsilon_{\mathrm{eff}} = \epsilon_1\epsilon_2,
	\qquad
	t_{\mathrm{eff}} = t_1t_2.
\end{equation}
\indent Accordingly, the probabilities of an erroneous conclusive
event and a conclusive event become
\begin{equation}
	p_{\mathrm{err}}^{\mathrm{pol}}
	= \frac{\epsilon_1\epsilon_2}{2}
	\label{eq:p-err-pol}
\end{equation}
and
\begin{equation}
	p_{\mathrm{det}}^{\mathrm{pol}}
	= \frac{t_1t_2+7\epsilon_1\epsilon_2}{8},
	\label{eq:p-det-pol}
\end{equation}
respectively.\\
\indent Let $\eta_t$ and $\eta_d$ denote the channel transmissivity
and detector efficiency, respectively, and let
$Y_0^{\mathrm{pol}}$ denote the background yield associated
with the conclusive detectors. Neglecting the simultaneous
occurrence of signal-induced and background-induced
detection events, the conclusive single-photon yield
can be expressed as
\begin{equation}
	Q_{\mathrm{pol}}
	= Y_0^{\mathrm{pol}}
	+ p_{\mathrm{det}}^{\mathrm{pol}}\eta_t\eta_d.
	\label{eq:gain-pol}
\end{equation}
\indent The quantum bit error rate (QBER) is then given by
\begin{equation}
	e_{\mathrm{pol}}
	=
	\frac{
		e_0Y_0^{\mathrm{pol}}
		+p_{\mathrm{err}}^{\mathrm{pol}}\eta_t\eta_d
	}{
		Q_{\mathrm{pol}}
	},
	\label{eq:qber-pol-general}
\end{equation}
where $e_0=1/2$ accounts for the random nature of
background-induced errors. Note that the last formula is adapted from Ref. \cite{ma2005}.\\
\indent Substituting equations~\eqref{eq:p-err-pol}--\eqref{eq:gain-pol}
into equation~\eqref{eq:qber-pol-general}, one obtains
\begin{equation}
	e_{\mathrm{pol}}
	=
	\frac{
		e_0Y_0^{\mathrm{pol}}
		+\frac{1}{2}\epsilon_1\epsilon_2\eta_t\eta_d
	}{
		Y_0^{\mathrm{pol}}
		+\frac{1}{8}
		(t_1t_2+7\epsilon_1\epsilon_2)\eta_t\eta_d
	}.
	\label{eq:qber-pol}
\end{equation}
\indent In the absence of background-induced detection events,
equation~\eqref{eq:qber-pol} reduces to
\begin{equation}
	e_{\mathrm{pol}}
	=
	\frac{
		4\epsilon_1\epsilon_2
	}{
		t_1t_2+7\epsilon_1\epsilon_2
	}.
\end{equation}
For $t_1,t_2\simeq 1$ and $\epsilon_1,\epsilon_2\ll 1$,
the latter expression can be approximated as
\begin{equation}
	e_{\mathrm{pol}}
	\simeq 4\epsilon_1\epsilon_2.
\end{equation}
\indent In particular, for two identical polarization analyzers
characterized by a leakage probability $\epsilon$, the
contribution of imperfect polarization discrimination
to the QBER scales quadratically with $\epsilon$, rather
than linearly, as in the case of a single PBS.
This improvement, however, comes at the expense of the
additional insertion loss introduced by the second
polarization analyzer.

\section{Error rate in phaseRQKD}
\label{er-phase}
Unlike polarization-encoded RQKD, phase-encoded RQKD is
inherently sensitive to fluctuations of the relative
phase between the interfering optical fields.
Accordingly, in addition to background-induced errors
and errors arising from imperfections of the optical
system, the contribution of turbulence-induced phase
fluctuations must be taken into account.\\
\indent Let $\sigma_\phi^2$ characterize the strength of the
turbulence-induced phase fluctuations. According to
the interference model considered herein, the probability
of an erroneous conclusive event arising exclusively
from atmospheric turbulence is given by
\begin{equation}
	p_{\mathrm{turb}}
	= \frac{1}{4}
	\left(1-\exp(-\sigma_\phi^2)\right),
	\label{eq:p-turb}
\end{equation}
whereas the overall probability of a conclusive event
amounts to
\begin{equation}
	p_{\mathrm{det}}^{\mathrm{ph}}
	= \frac{1}{2}
	- \frac{1}{4}\exp(-\sigma_\phi^2).
	\label{eq:p-det-phase}
\end{equation}
\indent Notice that Eq.~\eqref{eq:p-turb} describes an unconditional
probability of an erroneous conclusive event rather than
the corresponding QBER. The conditional error probability
associated with atmospheric turbulence is thus given by
\begin{equation}
	e_{\mathrm{turb}}
	=
	\frac{p_{\mathrm{turb}}}
	{p_{\mathrm{det}}^{\mathrm{ph}}}.
	\label{eq:e-turb}
\end{equation}

Let $e_{\mathrm{ph}}$ characterize the probability of an
error arising from imperfections of the optical system,
conditioned on the occurrence of a conclusive signal
detection event. Following Ma \textit{et al.}~\cite{ma2005},
this parameter is assumed to be independent of the
transmission distance.\\
\indent Since errors arising from imperfections of the optical
system and those induced by atmospheric turbulence may
occur simultaneously, their respective probabilities
cannot, in general, be added directly. Assuming that
the two error mechanisms act as statistically independent
binary reversals without altering the conclusive status
of a detection event, their simultaneous occurrence
restores the correct bit value.\\
\indent Accordingly, the conditional error probability associated
with signal-induced detection events is given by
\begin{equation}
	\begin{split}
		e_{\mathrm{sig}}^{\mathrm{ph}}
		&=
		e_{\mathrm{ph}}(1-e_{\mathrm{turb}})
		+(1-e_{\mathrm{ph}})e_{\mathrm{turb}}
		\\
		&=
		e_{\mathrm{ph}}+e_{\mathrm{turb}}
		-2e_{\mathrm{ph}}e_{\mathrm{turb}}.
	\end{split}
	\label{eq:e-sig-phase}
\end{equation}
\indent The unconditional probability of an erroneous conclusive
event is consequently given by
\begin{equation}
	p_{\mathrm{err}}^{\mathrm{ph}}
	=
	p_{\mathrm{det}}^{\mathrm{ph}}
	e_{\mathrm{sig}}^{\mathrm{ph}}.
\end{equation}
\indent Substituting equations~\eqref{eq:e-turb} and
\eqref{eq:e-sig-phase} into the latter expression yields
\begin{equation}
	p_{\mathrm{err}}^{\mathrm{ph}}
	=
	e_{\mathrm{ph}}p_{\mathrm{det}}^{\mathrm{ph}}
	+(1-2e_{\mathrm{ph}})p_{\mathrm{turb}}.
	\label{eq:p-err-phase}
\end{equation}
\indent Let $Y_0^{\mathrm{ph}}$ denote the background yield
associated with the conclusive detector of the
phase-encoded RQKD receiver. Neglecting the simultaneous
occurrence of signal-induced and background-induced
detection events, the conclusive single-photon yield
amounts to
\begin{equation}
	Q_{\mathrm{ph}}
	=
	Y_0^{\mathrm{ph}}
	+p_{\mathrm{det}}^{\mathrm{ph}}\eta_t\eta_d.
	\label{eq:gain-phase}
\end{equation}
\indent In accordance with the error-rate model of
Ma \textit{et al.}~\cite{ma2005}, adapted to the postselection
mechanism of phase-encoded RQKD \cite{bebrov,kravtsov}, the QBER is given by
\begin{equation}
	e_{\mathrm{phase}}
	=
	\frac{
		e_0Y_0^{\mathrm{ph}}
		+p_{\mathrm{err}}^{\mathrm{ph}}\eta_t\eta_d
	}{
		Q_{\mathrm{ph}}
	}.
\end{equation}
\indent Taking into account equations~\eqref{eq:p-err-phase} and
\eqref{eq:gain-phase}, one obtains
\begin{equation}
	e_{\mathrm{phase}}
	=
	\frac{
		e_0Y_0^{\mathrm{ph}}
		+
		\left[
		e_{\mathrm{ph}}p_{\mathrm{det}}^{\mathrm{ph}}
		+(1-2e_{\mathrm{ph}})p_{\mathrm{turb}}
		\right]\eta_t\eta_d
	}{
		Y_0^{\mathrm{ph}}
		+p_{\mathrm{det}}^{\mathrm{ph}}\eta_t\eta_d
	}.
	\label{eq:qber-phase}
\end{equation}
\indent In the absence of background-induced detection events,
equation~\eqref{eq:qber-phase} reduces to
\begin{equation}
	e_{\mathrm{phase}}
	=
	e_{\mathrm{ph}}
	+(1-2e_{\mathrm{ph}})
	\frac{
		1-\exp(-\sigma_\phi^2)
	}{
		2-\exp(-\sigma_\phi^2)
	}.
\end{equation}
\indent It is readily seen that, in the absence of atmospheric
turbulence, the QBER reduces to $e_{\mathrm{ph}}$.
In contrast, even in the absence of other optical imperfections
($e_{\mathrm{ph}}=0$), atmospheric turbulence gives rise
to a nonzero QBER.

\section{Detection probability in polarizationRQKD}
\label{det-prob-pol}
Let $P_{\perp}$ and $P_{\parallel}$ denote the ideal
probabilities of detecting a photon at the conclusive
and complementary polarization output ports, respectively,
summed over both spatial outputs of the interferometer.
Owing to imperfect polarization discrimination, the
actual probability of a conclusive detection event is
given by
\begin{equation}
	P_{\mathrm{con}}
	= t_1 P_{\perp} + \epsilon_1 P_{\parallel},
\end{equation}
where the first term accounts for the desired polarization
component, whereas the second term accounts for the
leakage of the undesired polarization component into
the conclusive output ports. \\
\indent For the polarization configurations $HH$ and $DD$,
one has $P_{\perp}=0$ and $P_{\parallel}=1$.
Consequently,
\begin{equation}
	P_{\mathrm{con}}(HH)
	= P_{\mathrm{con}}(DD)
	= \epsilon_1.
\end{equation}
All conclusive detection events associated with these
two configurations are erroneous.\\
\indent For the polarization configurations $HD$ and $DH$,
one has $P_{\perp}=1/4$ and $P_{\parallel}=3/4$.
Consequently,
\begin{equation}
	P_{\mathrm{con}}(HD)
	= P_{\mathrm{con}}(DH)
	= \frac{t_1}{4}+\frac{3\epsilon_1}{4}.
\end{equation}
Note that $t_1$ and $\epsilon_1$ are the transmission
probability associated with the desired output port and  the probability that a photon incident
on a polarization beam splitter (PBS) is routed to the
undesired output port (leakage probability), respectively (as mentioned in appendix \ref{er-pol})
Taking into account the equal a priori probabilities
of the four polarization configurations, the probability
of an erroneous conclusive event is given by
\begin{equation}
	p_{\mathrm{err}}^{\mathrm{pol}}
	= \frac{1}{4}(2\epsilon_1)
	= \frac{\epsilon_1}{2},
\end{equation}
whereas, according to the law of total probability,
the overall probability of a conclusive event amounts to
\begin{equation}
	\begin{split}
		p_{\mathrm{det}}^{\mathrm{pol}}
		&= \frac{1}{4}
		\left[
		2\epsilon_1
		+2\left(
		\frac{t_1}{4}
		+\frac{3\epsilon_1}{4}
		\right)
		\right]\\
		&= \frac{t_1+7\epsilon_1}{8}.
	\end{split}
\end{equation}

\end{document}